\pdfoutput=1
\documentclass{raa}

\usepackage{graphicx,times}
\usepackage{natbib}
\usepackage{amssymb,amsmath}
\usepackage{comment}
\usepackage{array}
\bibpunct{(}{)}{;}{a}{}{,}
\renewcommand{\figurename}{Fig.}
\renewcommand{\tablename}{Table}
\usepackage{hyperref}
\usepackage{soul}

\begin{document}

   \title{Reconstruction of Reionization Histories from 21\,cm Power-Spectrum Evolution with Artificial Neural Networks}

 \volnopage{ {\bf 20XX} Vol.\ {\bf X} No. {\bf XX}, 000--000}
   \setcounter{page}{1}

\author{Yu-Le Wang\inst{1,2*}, Hayato Shimabukuro\inst{1,2,3}
}
\institute{
South-Western Institute for Astronomy Research (SWIFAR), Yunnan University, Kunming, Yunnan 650500, China\\
\vs \no
Key Laboratory of Survey Science of Yunnan Province, Yunnan University, Kunming, Yunnan 650500, China\\
\vs \no
Graduate School of Science, Division of Particle and Astrophysical Science, Nagoya University, Chikusa-ku, Nagoya, 464-8602, Japan\\
\vs \no
{\small Received 20XX Month Day; accepted 20XX Month Day}
}

\vs \no
   {\small Received 20XX Month Day; accepted 20XX Month Day}

\email{1225960571@qq.com}

\abstract{We investigate whether the redshift evolution of the fixed-\(k\) dimensionless 21\,cm power spectrum, \(\Delta^2_{21}(k,z)\), contains sufficient information to reconstruct reionization histories \(x_{\mathrm{HI}}(z)\) with artificial neural networks. Using semi-numerical realizations generated within a restricted three-parameter 21cmFAST model family, we train a compact feed-forward network to learn the inverse mapping from power-spectrum trajectories to the neutral-fraction history on a 97-point redshift grid spanning \(6 \le z \le 15\). For \(k=0.1\), \(0.5\), and \(1.0~h\,\mathrm{Mpc}^{-1}\), representative tests on an independent test set show that the midpoint redshift \(z_{50}\) is recovered more accurately than the duration \(\Delta z=z_{75}-z_{25}\) in terms of dimensionless relative errors: \(z_{50}\) is reconstructed with \(\mathrm{MAE}=0.0046\) and \(\mathrm{RMSE}=0.0100\), whereas \(\Delta z\) yields \(\mathrm{MAE}=0.0302\) and \(\mathrm{RMSE}=0.0378\). This result indicates that fixed-\(k\) power-spectrum evolution carries stronger information about the timing of reionization than about the detailed width of the transition within the adopted prior. We further test an idealized foreground-free SKA1-Low-like thermal-plus-sample-variance noise model and find that the reconstruction remains stable in the favorable signal-to-noise regime considered here. These results demonstrate that neural networks can serve as prior-dependent inverse mapping for reconstructing reionization histories from 21\,cm power-spectrum evolution.
}

   \authorrunning{Wang \& Shimabukuro}            
   \titlerunning{Reionization Histories from 21\,cm Power-Spectrum Evolution}  
   \maketitle

%
\section{INTRODUCTION}           
\label{sect:intro}
The Cosmic Dawn and the Epoch of Reionization mark the interval during which the first luminous sources transformed the intergalactic medium from a predominantly neutral state into an ionized one. A central goal of early-Universe astrophysics is therefore to reconstruct the redshift evolution of the volume-averaged neutral hydrogen fraction, $x_{\rm HI}(z)$. That history encodes when the first galaxies became efficient ionizing sources, how rapidly ionized regions grew and merged, and how radiative feedback shaped subsequent structure formation. By $z \approx 6$, reionization was largely complete, but the timing and duration of the transition remain key open questions \citep[e.g.][]{2001PhR...349..125B,2022ARA&A..60..121R}.

Recent observations have made this question increasingly concrete. On the galaxy side, JWST spectroscopy is beginning to probe the luminosities, ionizing spectra, and local environments of sources well into the reionization era, while Ly$\alpha$-based measurements---including absorption-line constraints from Gunn--Peterson troughs and damping wings, as well as luminosity-function and clustering studies of Ly$\alpha$ emitters---provide increasingly informative, though model-dependent, constraints on the neutral IGM at $z \gtrsim 6$. Joint interpretations of these data have reinforced the view that the main phase of reionization may occur relatively late, with a substantial decline in $x_{\rm HI}$ between $z\sim 8$ and $z\sim 6$, and in some analyses even a comparatively sharp transition, although the inferred history is not unique \citep{2006ARA&A..44..415F,2018PhR...780....1D,2017ApJ...851...50M,2025arXiv250404683U,Umeda_2025,2025ApJS..278...33K}.

At the same time, these observables do not directly provide the full function $x_{\rm HI}(z)$. JWST primarily constrains the abundance, spectra, and ionizing efficiency of source populations, so converting those data into a global ionization history requires assumptions about escape fractions, source demographics, and the coupling between galaxies and the surrounding IGM \citep{2018PhR...780....1D,2017ApJ...851...50M}. Ly$\alpha$-based probes are especially informative during the late stages of reionization, but their interpretation depends on absorption-line radiative transfer and on the environments of the background sources: Gunn--Peterson and damping-wing measurements are sensitive to neutral hydrogen along particular sightlines, whereas Ly$\alpha$ emitter statistics respond to both IGM transmission and galaxy evolution. In addition, the Ly$\alpha$ forest saturates rapidly as the neutral fraction rises, which limits the direct leverage of absorption measurements on the full reionization history and increases the importance of astrophysical systematics associated with circumgalactic gas and large-scale ionized topology \citep{2006ARA&A..44..415F,2025arXiv250404683U,Umeda_2025,2025ApJS..278...33K}. The motivation for additional probes is therefore not that current measurements are inadequate, but that reconstructing $x_{\rm HI}(z)$ benefits from observables that are complementary to them and that trace neutral hydrogen in the IGM more directly.

In this context, the redshifted 21\,cm signal is especially attractive. Because the differential brightness temperature depends on the density field, the neutral fraction, and the spin temperature relative to the cosmic microwave background, the 21\,cm signal can in principle follow the evolution of the neutral IGM over a broad redshift range and remain sensitive to the growth, overlap, and patchiness of ionized regions \citep[e.g.,][]{1990MNRAS.247..510S, 1997ApJ...475..429M, 2004ApJ...613....1F, 2006PhR...433..181F, 2012RPPh...75h6901P,2023PASJ...75S...1S}. For interferometric experiments such as LOFAR, HERA, and SKA, the power spectrum has become the natural and most practical summary statistic, because it captures this evolving morphology in a form that can be measured before high-fidelity imaging becomes routine \citep{FurlanettoBriggs2004,Zaldarriaga2004,MoralesWyithe2010,2015aska.confE...1K,2017PASP..129d5001D}.

Even in the 21\,cm case, however, the scientific target of this paper, $x_{\rm HI}(z)$, is not measured directly. The 21 cm observable used here is the redshift evolution of the fixed-$k$ dimensionless power spectrum, \(\Delta^2_{21}(k,z)\), so the task is to infer a reionization-history trajectory from an observable power-spectrum trajectory. Within the adopted three-parameter 21cmFAST prior, this is a nonlinear inverse map rather than a direct measurement, and it is not obvious that the relation can be reduced to a simple analytic fitting formula. Forward-modeling frameworks based on 21cmFAST and related Bayesian inference pipelines provide an established way to constrain astrophysical parameters and average neutral fractions within explicit priors, but they are not the most direct tool for the proof-of-concept question posed here: whether this trajectory-to-trajectory inverse relation is sufficiently regular and informative to be learned within a restricted model family \citep{2011MNRAS.411..955M,2020JOSS....5.2582M,10.1093/mnras/stv571}.

This motivates the use of a compact ANN as a flexible inverse mapping. Its role in the present study is not to provide a model-independent estimator, but to test whether fixed-$k$ trajectories of \(\Delta^2_{21}(k,z)\) retain enough information to reconstruct $x_{\rm HI}(z)$ within a restricted three-parameter 21cmFAST prior. Machine-learning methods have already been explored in several 21\,cm contexts, including early ANN analyses of the 21\,cm signal, CNN-based inference of reionization history from 21\,cm images, ANN recovery of the H II bubble size distribution from the power spectrum, and nonlinear reconstruction of the 21\,cm global signal from power-spectrum information \citep[e.g.][]{2017MNRAS.468.3869S,La_Plante_2019,2022RAA....22c5027S,2024MNRAS.527.9977S,2025RAA....25h5017S}. Motivated by these works, we use the ANN here as a prior-dependent proof-of-concept tool and evaluate how well it reconstructs $x_{\rm HI}(z)$ 
from simulated trajectories spanning $z=6$--$15$, both in the noiseless case and under an idealized foreground-free SKA1-Low-like thermal-plus-sample-variance noise model. The remainder of this paper introduces the 21\,cm observables and target histories, describes the ANN framework, and then presents representative results for both noiseless and noisy cases.

\section{21cm Signal, Power Spectrum, and Target Reionization History}
\label{sec:background}

This section summarizes the physical quantities used throughout this work and clarifies the
inverse problem addressed in the following sections. We first review the basic physics of the
cosmological 21\,cm signal, then define the reionization history \(x_{\mathrm{HI}}(z)\) as the
target quantity to be reconstructed, and finally introduce the power spectrum as the
interferometric observable used to infer it.

\subsection{21\,cm brightness temperature}
\label{subsec:dtb}

The redshifted 21\,cm hyperfine transition of neutral hydrogen is one of the most promising probes
of the intergalactic medium during cosmic dawn and the epoch of reionization
\citep{1990MNRAS.247..510S,1997ApJ...475..429M,2006PhR...433..181F,2012RPPh...75h6901P}.
Its observability depends on the spin temperature \(T_S\), which is regulated by radiative coupling
to the CMB, collisional processes, and Ly\(\alpha\) pumping through the Wouthuysen--Field mechanism
\citep{Wouthuysen1952,Field1958,Field1959,Hirata2006,ChenMiralda2004,ChenMiralda2008}.
In the fiducial simulations used in this work, spin-temperature fluctuations are explicitly included.
The associated X-ray heating parameters are fixed to the default 21cmFAST values and are not varied
within the training prior. This is a deliberate simplification: for the present proof-of-concept we
vary only the parameters that most directly control the timing and morphology of reionization, so
that the network learns a restricted inverse map to \(x_{\mathrm{HI}}(z)\) rather than a broader joint
heating-ionization problem. Accordingly, the factor \(1-T_\gamma/T_S\) is computed self-consistently
in the forward model rather than being assumed to be unity over the redshift range used for the
reconstruction.

For a gas element at redshift \(z\) and comoving position \(\boldsymbol{x}\), the differential 21\,cm
brightness temperature relative to the CMB can be approximated as \citep{2019cosm.book.....M}
\begin{equation}
\delta T_b(z,\boldsymbol{x}) \cong
27(1+\delta)\,x_{\mathrm{HI}}(z,\boldsymbol{x})
\left(\frac{\Omega_{b,0} h^2}{0.023}\right)
\left(\frac{0.15}{\Omega_{m,0} h^2}\frac{1+z}{10}\right)^{1/2}
\left(1-\frac{T_\gamma}{T_S}\right)\ {\rm mK},
\label{eq:dtb}
\end{equation}
where \(\delta\) is the baryon overdensity, \(x_{\mathrm{HI}}(z,\boldsymbol{x})\) is the local neutral
hydrogen fraction, \(T_\gamma\) is the CMB temperature, and \(T_S\) is the hydrogen spin
temperature. The corresponding local ionized fraction is \(1-x_{\mathrm{HI}}(z,\boldsymbol{x})\).
For simplicity, we use the approximate form above and do not explicitly include the peculiar-velocity
gradient term, since the purpose of this paper is not precision forward modeling of the full 21\,cm
signal but the inverse reconstruction of the global reionization history.

Equation~(\ref{eq:dtb}) makes clear that the 21\,cm signal is intrinsically multivariate: it depends
on the density, ionization, and thermal states of the gas, as well as on the radiation background.
The mapping from a 21\,cm statistic back to the global reionization history is therefore nonlinear
and prior dependent. Throughout this work we assume a flat \(\Lambda\)CDM cosmology consistent
with \textit{Planck} 2018, with \(\Omega_m = 0.315\), \(\Omega_b = 0.049\), \(h = 0.674\),
\(\sigma_8 = 0.811\), and \(n_s = 0.965\) \citep{2020A&A...641A...6P}.

\subsection{Target reionization history}
\label{subsec:history}

The quantity reconstructed in this paper is the volume-averaged neutral fraction
\(x_{\mathrm{HI}}(z)\), which we refer to as the reionization history. When convenient, we also use
the ionized volume filling factor
\begin{equation}
Q(z) = 1 - x_{\mathrm{HI}}(z).
\label{eq:Qdef}
\end{equation}
This is the natural target quantity for the present study, because it provides a compact description
of the global progress of reionization and allows summary quantities such as the midpoint and
duration to be defined straightforwardly.

A useful approximate description of the global reionization history is given by the photon-counting
equation \citep{2017ApJ...851...50M}
\begin{equation}
\frac{dQ}{dt}
=
\frac{\langle \dot{n}_{\mathrm{ion}} \rangle}{\langle n_{\mathrm{H}} \rangle}
-
\frac{Q}{\bar{t}_{\mathrm{rec}}},
\label{eq:reion}
\end{equation}
where \(\langle \dot{n}_{\mathrm{ion}} \rangle\) is the emission rate of ionizing photons escaping into the
IGM per unit proper volume, \(\langle n_{\mathrm{H}} \rangle = 1.89 \times 10^{-7}(1+z)^3~\mathrm{cm}^{-3}\)
is the cosmological mean hydrogen number density, and \(\bar{t}_{\mathrm{rec}}\) is an effective
recombination timescale \citep{1999ApJ...514..648M}. Equation~(\ref{eq:reion}) makes clear
that reionization histories depend on both source efficiency and recombinations, so any inversion
from 21\,cm statistics necessarily remains sensitive to the adopted astrophysical prior.

Since the midpoint and duration of reionization play an important role in the analysis below,
we define them explicitly here. Here, the ``midpoint'' simply means the epoch at which
reionization is halfway complete, so that the IGM is 50\% neutral on average. The midpoint
redshift \(z_{50}\) is defined by
\begin{equation}
x_{\mathrm{HI}}(z_{50}) = 0.5,
\label{eq:z50def}
\end{equation}
and the reionization duration is defined as
\begin{equation}
\Delta z \equiv z_{75} - z_{25},
\label{eq:deltazdef}
\end{equation}
where \(x_{\mathrm{HI}}(z_{75}) = 0.75\) and \(x_{\mathrm{HI}}(z_{25}) = 0.25\). We adopt these
thresholds because they bracket the central 50\% of the transition, providing a simple measure
of its width while avoiding excessive sensitivity to the slowly varying early and late tails of
the history near \(x_{\mathrm{HI}} \approx 1\) and \(x_{\mathrm{HI}} \approx 0\). With this convention,
\(\Delta z\) is always positive and quantifies the redshift width of the main reionization
transition.

Recent JWST-based analyses suggest that reionization may have been relatively late and rapid,
with a steep decline in \(x_{\mathrm{HI}}\) toward \(z \sim 6\)--8
\citep{2025ApJS..278...33K,2025arXiv250404683U,Umeda_2025}. We use these studies only to
motivate the compressed parameter ranges explored below, not as a claim that the true reionization
history is uniquely determined.

\begin{figure}[ht]
\centering
\includegraphics[width=14cm]{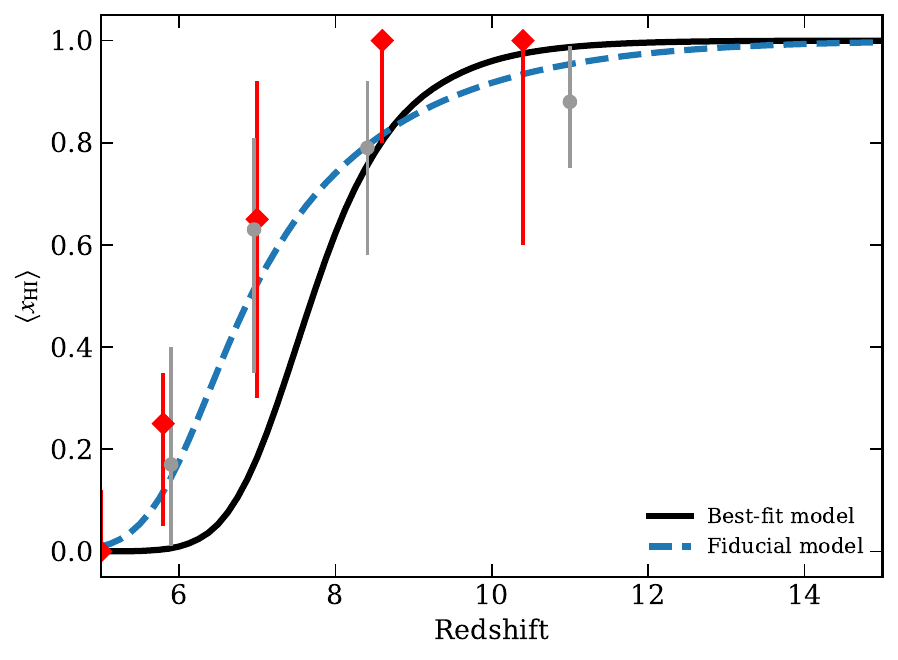}
\caption{Comparison between a more traditional reionization history and a late-reionization history
motivated by recent JWST-based analyses \citep{Umeda_2025,2025ApJS..278...33K}.
The black solid curve shows a best-fit late-reionization model, while the blue dashed curve shows
a fiducial faint-galaxy-dominated history. Symbols denote observationally inferred neutral fractions
from the cited studies.}
\label{fig:1}
\end{figure}

Figure~\ref{fig:1} illustrates this motivation by comparing a more traditional reionization history
with a late-reionization history inspired by recent observational analyses. The purpose of this
comparison is only to indicate the type of target histories emphasized in the present work.

\subsection{21\,cm power spectrum as the observable}
\label{subsec:ps}

For interferometric observations, the 21\,cm power spectrum is the natural summary statistic,
because radio arrays measure Fourier modes of the sky brightness rather than the sky-averaged
signal directly \citep{FurlanettoBriggs2004,Zaldarriaga2004,MoralesWyithe2010,2015aska.confE...1K}.
Following \citet{2006PhR...433..181F} and \citet{2019cosm.book.....M}, we define the brightness-temperature fluctuation field as
\begin{equation}
\delta T_{21}(\mathbf{x})
= \delta T_b(\mathbf{x})-\langle \delta T_b \rangle ,
\end{equation}
and its power spectrum by
\begin{equation}
\left\langle 
\delta T_{21}(\mathbf{k}_1)
\delta T_{21}^{*}(\mathbf{k}_2)
\right\rangle
=
(2\pi)^3 \delta_{\rm D}(\mathbf{k}_1-\mathbf{k}_2)
P_{21}(k_1).
\end{equation}
The dimensionless power spectrum is then
\begin{equation}
\Delta^2_{21}(k)
=
\frac{k^3}{2\pi^2}P_{21}(k),
\end{equation}
which has units of ${\rm mK}^2$.

The 21\,cm power spectrum is sensitive to the evolving topology of ionized regions, the global
neutral fraction, and the thermal state of the IGM. Its scale dependence is therefore physically
informative: on relatively large scales, the signal is more directly related to the growth and overlap
of ionized bubbles, whereas on smaller scales it becomes increasingly sensitive to local morphology
and substructure \citep{Zaldarriaga2004,2006PhR...433..181F,MoralesWyithe2010,2015aska.confE...1K}.

In this work, we focus on the redshift evolution of the dimensionless 21\,cm power spectrum at fixed
wavenumber, denoted by \(\Delta^2_{21}(k,z)\). The scale dependence of this statistic reflects the
characteristic sizes of ionized regions and the evolving contrast between neutral and ionized gas.
On relatively large scales, the power spectrum is more directly linked to the global progress of
reionization, whereas on smaller scales it becomes increasingly sensitive to local morphology and
substructure \citep{Zaldarriaga2004,2006PhR...433..181F,MoralesWyithe2010,2015aska.confE...1K}.
This scale dependence is important for interpreting how different Fourier modes trace the evolving
ionization structure of the intergalactic medium.

Equation~(\ref{eq:dtb}) also provides the physical reason that the redshift evolution of the fixed-$k$ power spectrum can encode information about the EoR history. The brightness-temperature field is modulated by the local neutral fraction, density fluctuations, and spin-temperature factor. After subtraction of the mean brightness temperature, the power spectrum measures the variance of the remaining spatial fluctuations, so changes in the abundance and sizes of neutral and ionized regions leave a redshift-dependent imprint on $P_{21}(k)$. During the early stages, most of the IGM is neutral and the contrast is dominated mainly by density and temperature structure; as ionized bubbles grow and merge, the contrast between ionized and neutral regions transfers power to scales comparable to the characteristic bubble size; toward the end of reionization, the shrinking neutral volume suppresses the 21 cm signal. Thus a fixed-$k$ trajectory is not an arbitrary time series, but a compressed record of how the ionization topology passes through the corresponding spatial scale \citep{Zaldarriaga2004,2004ApJ...613....1F,2006PhR...433..181F,MoralesWyithe2010}. In the restricted simulations used here, the heating sector is held fixed, which further reduces the dimensionality of this relation and makes the power-spectrum evolution more directly tied to the ionization history than it would be in a fully general model. This physical connection motivates using the trajectory \(\Delta^2_{21}(k,z)\rightarrow x_{\mathrm{HI}}(z)\) as the inverse mapping tested in this proof-of-concept study, while also explaining why the mapping remains scale dependent and prior dependent.

\begin{figure}[!htbp]
\centering
\includegraphics[width=14cm]{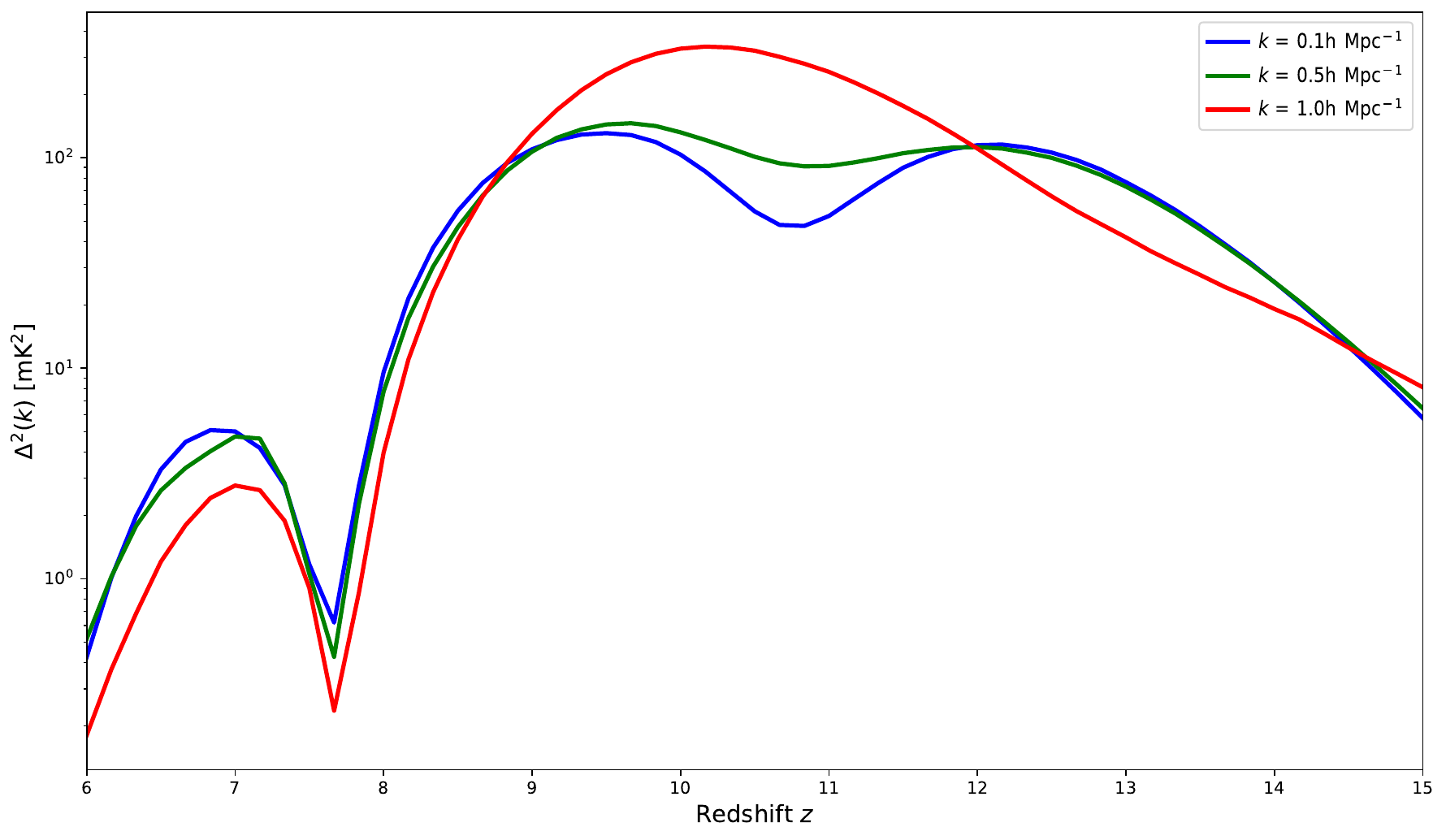}
\caption{Redshift evolution of the dimensionless power spectrum \(\Delta^2_{21}(k,z)\) for the fixed-$k$ trajectories considered in this work.}
\label{fig:6}
\end{figure}

Figure~\ref{fig:6} shows the redshift evolution of
\(\Delta^2_{21}(k,z)\) for the fixed-$k$ trajectories considered in this work.

With the target quantity \(x_{\mathrm{HI}}(z)\) defined above and the observable
\(\Delta^2_{21}(k,z)\) introduced here, the present work studies the inverse mapping
\begin{equation}
\Delta^2_{21}(k,z)\ \longrightarrow\ x_{\mathrm{HI}}(z).
\end{equation}
We do not attempt a model-independent reconstruction; instead, this inverse relation is examined
within a restricted but physically motivated three-parameter 21cmFAST model family.

\section{ARTIFICIAL NEURAL NETWORK}
The inverse problem considered in this work is to reconstruct the reionization history
\(x_{\mathrm{HI}}(z)\) from the redshift evolution of the dimensionless 21\,cm power spectrum at fixed wavenumber.
The central question is, therefore, how accurately the reionization history can be reconstructed
from a fixed-\(k\) 21\,cm power-spectrum trajectory for the restricted set of histories generated
by the three-parameter 21cmFAST setup adopted here.

We use a feed forward ANN as a mapping that maps the redshift evolution of \(\Delta^2_{21}(k,z)\) at fixed $k$ to the target reionization history. The point of this section is therefore not to review generic deep learning concepts, but to describe the specific mapping used to test whether the inverse map is sufficiently regular within this restricted three-parameter 21cmFAST setup. By the universal approximation theorem, even a network with a single hidden layer can represent a wide class of smooth nonlinear maps \citep{HORNIK1989359}, which is adequate for the restricted problem considered here.

\subsection{Data Preparation}
We generate the dataset with 21cmFAST v3.4 \citep{2011MNRAS.411..955M,2019MNRAS.484..933P,2020JOSS....5.2582M}, which uses the excursion set formalism \citep{2004ApJ...613....1F} to evolve the density and ionization fields and compute the corresponding 21cm signal. We vary only three astrophysical parameters and sample them with Latin hypercube sampling (LHS), following the parameterization widely used in 21cmFAST-based inference studies \citep{10.1093/mnras/stv571}:

(1) \texttt{HII\_EFF\_FACTOR}($\zeta$): the ionizing efficiency of high $z$ galaxies. Higher values tend to speed up reionization.

(2) \texttt{ION\_Tvir\_MIN}($T_{\mathrm{vir}}$): minimum virial temperature of star forming haloes, given in $\log_{10}$ units.

(3) \texttt{R\_BUBBLE\_MAX}($R_{\mathrm{mfp}}$): Mean free path in Mpc of ionizing photons within ionized regions.

Motivated by the late reionization scenarios discussed in Section 2.3, we restrict the parameter ranges to those listed in Table~\ref{tab:params}. This compression is deliberate: it lets us ask how much information fixed $k$ power-spectrum evolution carries inside a narrow but astrophysically motivated family of histories. In particular, we vary only the three parameters that most directly regulate the ionization-side reionization history in this setup---\texttt{HII\_EFF\_FACTOR}, \texttt{ION\_Tvir\_MIN}, and \texttt{R\_BUBBLE\_MAX}---while keeping the X-ray heating sector fixed at the default 21cmFAST values. The aim is to isolate a prior-dependent inverse problem for $x_{\mathrm{HI}}(z)$: if X-ray heating parameters were also varied, \(\Delta^2_{21}(k,z)\) would respond to both thermal and ionization histories, broadening the allowed trajectory family and introducing additional heating-ionization degeneracies that are beyond the scope of this proof-of-concept. Consequently, the analysis does not test whether the network can disentangle a broad heating-ionization degeneracy; instead it probes a restricted inverse problem conditioned on an ionization-side prior.
\begin{table}[htbp]
\begin{center}
\caption[]{Parameter Settings for the Semi Numerical Reionization History Simulation}\label{tab:params}


 \begin{tabular}{cllc}
  \hline\noalign{\smallskip}
Symbol  &  Parameter Description      & 21cmFAST Parameter & Value Range                    \\
  \hline\noalign{\smallskip}
$\zeta$  & Ionizing Efficiency & \texttt{HII\_EFF\_FACTOR}     & $[260, 400]$  \\
$\log T_{\rm vir}^{\rm min}$     &   Min. Virial Temperature (log K)     &   \texttt{ION\_Tvir\_MIN}     & $[5.45, 5.85]$                  \\
$R_{\rm mfp}$     &   Max. Mean Free Path (Mpc)     &   \texttt{R\_BUBBLE\_MAX}     & $[10, 65]$                  \\
  \noalign{\smallskip}\hline
\end{tabular}
\end{center}
\end{table}





We emphasize that the compressed parameter ranges listed in Table 1 are motivated by recent JWST-based analyses suggesting relatively late and rapid reionization scenarios (e.g., \mbox{\citep{Umeda_2025,2025ApJS..278...33K}}). We use these ranges to bracket observationally motivated parameter values and to define a controlled, compact prior for this proof-of-concept test. They are not intended to cover the full space of possible EoR histories or to identify the true cosmic reionization history.

In the implementation scripts, each simulated realization contributes one fixed $k$ trajectory sampled at 97 linearly spaced redshift bins spanning $z=6$--$15$. We generate 200 realizations for each of the three fixed wavenumbers considered here ($k=0.1$, $0.5$, and $1.0~h\,\mathrm{Mpc}^{-1}$), for a total of 600 trajectories. Accordingly, the ANN input and output vectors both have dimension 97, and all 97 bins are used during training and prediction. For the noisy recovery pipeline supplied with the paper, the explicit simulation and noise generation script is written for the representative case $k_{\rm target}=0.1~h\,\mathrm{Mpc}^{-1}$, while the same 97-dimensional architecture is used for the scale-dependent tests shown in the manuscript.

Because each fixed-$k$ dataset contains only 200 trajectories, the numerical experiment should be interpreted as a representative in-prior proof of concept rather than as a final assessment of out-of-prior generalization. The sampled input and output vectors are 97-dimensional, but their intrinsic variability is strongly constrained by the three varied astrophysical parameters and by the smoothness of the resulting reionization histories and power-spectrum trajectories. This low-dimensional prior structure makes the present exercise different from fitting arbitrary high-dimensional noisy data, although it does not remove the risk of overfitting. For each fixed $k$, we therefore use a leakage-safe train/validation/test partition, with the validation set used for early stopping and checkpoint selection and the held-out test set used only for final evaluation. The reported metrics are representative one-split results for the adopted simulation prior, not fold-averaged population-level errors.

The simulation script uses \texttt{HII\_DIM}=200 and \texttt{BOX\_LEN}=300. 
The raw power spectrum is computed from the 21\,cm brightness temperature 
cube with 100 linearly spaced $k$ bins between $0.05$ and $1.5~h\,\mathrm{Mpc}^{-1}$, and the nearest bin to the target $k$ is converted 
to the dimensionless form $\Delta^2_{21}(k)$. All ANN inputs discussed below 
are these fixed-$k$ \(\Delta^2_{21}(k,z)\) trajectories.

\subsection{Artificial Neural Network Architecture}
Our ANN is a standard multilayer perceptron with one hidden layer. 
In the implementation used here, the network maps a 97-dimensional fixed-$k$ 
$\Delta^2_{21}(k,z)$ trajectory to a 97-dimensional reconstructed reionization 
history. The architecture consists of an input layer with 97 nodes, one hidden 
layer with 128 neurons, a ReLU activation function, and an output layer with 
97 nodes. In code, this corresponds to the sequence
\texttt{Linear(97,128)} $\rightarrow$ ReLU $\rightarrow$ 
\texttt{Linear(128,97)}.

This architecture contains two dense weight matrices with $97\times128+128\times97=24{,}832$ weights, or $25{,}057$ trainable parameters when biases are included. This parameter count is large compared with the number of training trajectories in each fixed-$k$ split, so overfitting is a real concern. We therefore do not interpret the network complexity as evidence of broad predictive power; instead, it is used only to test whether a smooth inverse map can be learned within the restricted three-parameter trajectory family.

Let $x_j$ denote the input power-spectrum trajectory sampled on the redshift grid. 
The pre-activation of the hidden layer is
\begin{equation}
\label{eq:hidden}
s_i = \sum_{j=1}^{N_{\rm in}} \omega_{ij}^{(1)} x_j + b_i^{(1)},
\end{equation}
where $N_{\rm in}=97$, $\omega_{ij}^{(1)}$ are the first-layer weights, and 
$b_i^{(1)}$ are the corresponding biases. The activated hidden representation is
\begin{equation}
t_i = \mathrm{ReLU}(s_i).
\end{equation}
The output layer then gives
\begin{equation}
\label{eq:output}
y_i = \sum_{j=1}^{N_{\rm h}} \omega_{ij}^{(2)} t_j + b_i^{(2)},
\end{equation}
where $N_{\rm h}=128$, and $y_i$ denotes the reconstructed neutral fraction at 
the $i$-th redshift bin.

The role of the ANN is therefore to learn a nonlinear map from 
$\Delta^2_{21}(k,z)$ to $x_{\mathrm{HI}}(z)$ within the adopted training family. 
We do not interpret the network as discovering a model-independent reionization 
history from first principles.

\subsection{Back Propagation Algorithm}
We optimize the network weights with standard back propagation \citep{1986Natur.323..533R}. The total training cost is
\begin{equation}
\label{eq:cost}
E = \sum_{n=1}^{N_{\text{train}}} E_n = \sum_{n=1}^{N_{\text{train}}} \left[ \frac{1}{2} \sum_{i=1}^{m} (y_{i,n} - d_{i,n})^2 \right]
\end{equation}

where $N_{\text{train}}$ is the number of training samples, $m$ is the number of output neurons, $y$ is the ANN prediction, and $d$ is the target. The weight update follows
\begin{equation}
\label{eq:update}
\Delta \omega_{i,j}^{(l)} = -\eta \frac{\partial E}{\partial \omega_{i,j}^{(l)}} = -\eta \sum_{n=1}^{N_{\text{train}}} \frac{\partial E}{\partial \omega_{i,j}^{(l)}}
\end{equation}

where $\eta$ is the learning rate. For the purposes of this paper, the technical role of training is straightforward: it adjusts the inverse-mapping so that the predicted $x_{\mathrm{HI}}(z)$ tracks the simulation target on unseen samples from the same prior family.

\subsection{Experimental Settings}
We use the rectified linear unit (ReLU) as the hidden layer activation function \citep{2010Rectified},
\begin{equation}
\label{eq:relu}
f(x) = \max(0, x)
\end{equation}

and the mean squared error (MSE) as the training loss,
\begin{equation}
\label{eq:mse}
\text{MSE} = \frac{1}{N_{\text{train}}} \sum_{i=1}^{N_{\text{train}}} E_n
\end{equation}

The training process is through PyTorch and run on either GPU or CPU depending on availability. The clean signal recovery uses the Adam optimizer with learning rate $10^{-2}$, a maximum of 2000 epochs, and early stopping with patience 100 based on the validation loss. The thermal noise recovery keeps the same ANN architecture but switches to AdamW with learning rate $10^{-3}$ and weight decay $10^{-4}$, using up to 10,000 epochs. In that script the SKA-like noise case uses patience 100, while the exaggerated $1000\times$ noise case uses patience 500 to tolerate larger validation loss fluctuations. In both scripts, the best checkpoint is chosen solely from the validation loss and then applied to the test set. All preprocessing transformations are fitted on the training subset only and then applied to the validation and test subsets. This procedure cannot remove the fundamental limitation imposed by the small simulation set, but it reduces data leakage and prevents the test samples from influencing model selection.

\begin{figure}[ht]
\centering
\includegraphics[width=14cm]{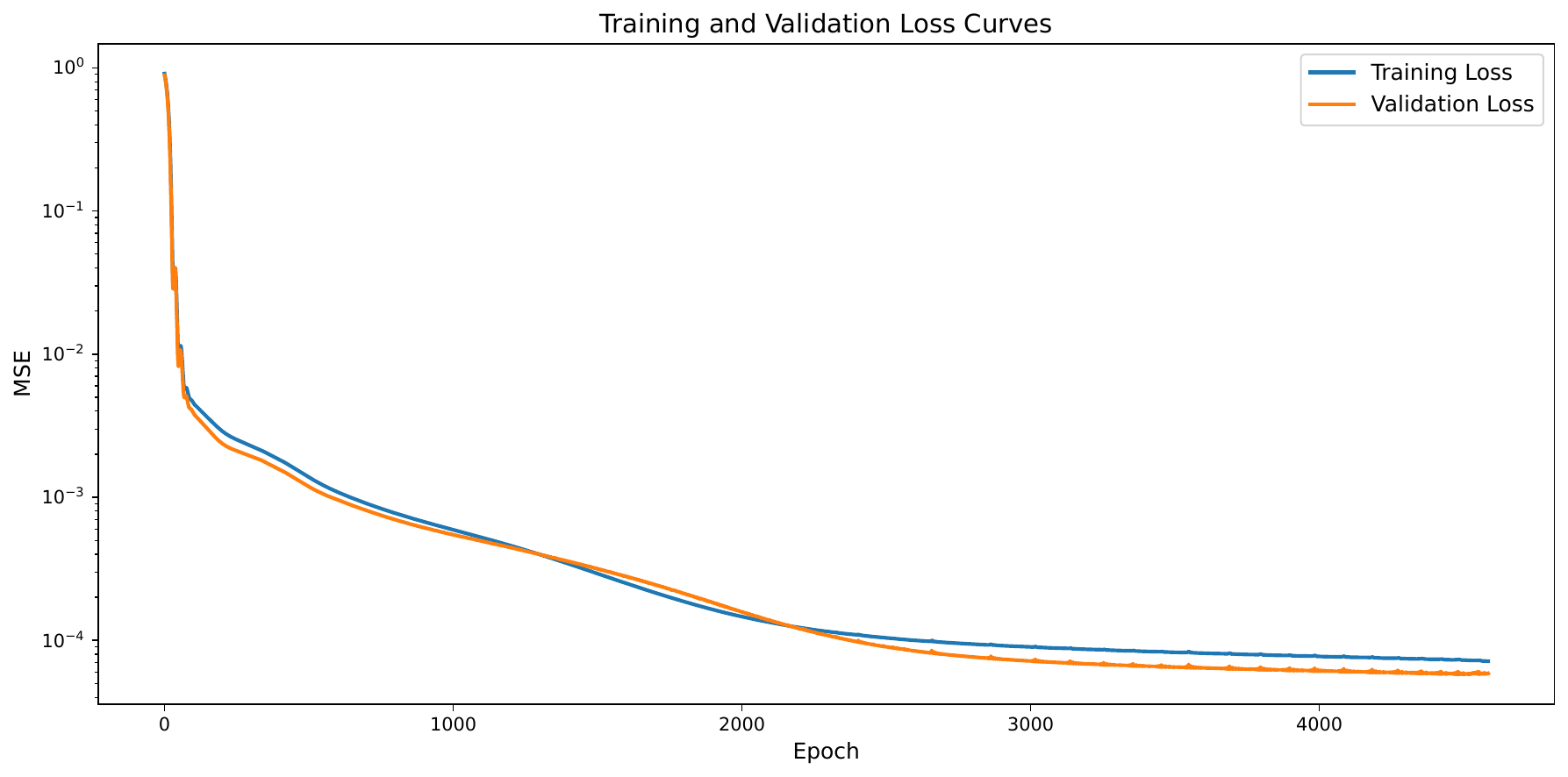}
\caption{Representative MSE learning curves during model optimization. Early stopping is applied using the validation set MSE, and the best validation checkpoint is retained for final evaluation.}
\label{fig:3}
\end{figure}

Figure~\ref{fig:3} shows representative learning curves. In the representative run shown in Fig. 3, the training and validation losses decrease together to the approximately $10^{-4}$ level, without a visible late-epoch separation. This behavior is a useful internal check against catastrophic in-prior overfitting, but it should not be interpreted as proof of out-of-prior generalization. In this study, training, validation, and diagnostic test losses are all recorded, but only the validation loss is used for checkpoint selection. We evaluate three aspects of the inverse problem on one independent test split. First, we compare reconstructions at fixed $k=0.1$, $0.5$, and $1.0~h\,\mathrm{Mpc}^{-1}$ to test how the information content changes with scale. Second, we add an idealized foreground-free SKA1-Low-like thermal-plus-sample-variance noise model, as well as an exaggerated $1000\times$ noise stress test, to assess how stable the inverse map remains in noisier conditions. Third, because the restricted model family produces smooth S-shaped histories, we examine the transition interval $0.2 \le x_{\mathrm{HI}} \le 0.8$ as a scientifically motivated test of whether the network is capturing the physically informative part of reionization rather than only the generic endpoints.

As an additional capacity check, we trained the same one-hidden-layer network on the existing representative $k=0.1~h\,\mathrm{Mpc}^{-1}$ dataset while varying only the hidden-layer width. We used hidden sizes 16, 32, 64, and 128, five random initialization seeds for each width, the same train/validation/test split, AdamW with learning rate $10^{-3}$ and weight decay $10^{-4}$, and validation-based early stopping. This test is not a substitute for a larger simulation suite, but it checks whether the result is an obvious artifact of one over-parameterized 128-neuron model.

\begin{table}[!htbp]
\begin{center}
\caption[]{Hidden-size ablation on the existing representative $k=0.1~h\,\mathrm{Mpc}^{-1}$ dataset. Values are mean $\pm$ standard deviation over five random seeds. For $x_{\rm HI}$, entries are absolute neutral-fraction MAE over the transition interval; for $z_{50}$ and $\Delta z$, entries are dimensionless relative MAE.\label{tab:hidden_ablation}}
\begin{tabular}{cccccc}
\hline\noalign{\smallskip}
Hidden size & Parameters & Seeds & $x_{\rm HI}$ MAE & $z_{50}$ MAE & $\Delta z$ MAE \\
\hline\noalign{\smallskip}
16  & 3217  & 5 & 0.0402 $\pm$ 0.0052 & 0.0120 $\pm$ 0.0011 & 0.0587 $\pm$ 0.0036 \\
32  & 6337  & 5 & 0.0418 $\pm$ 0.0061 & 0.0130 $\pm$ 0.0020 & 0.0716 $\pm$ 0.0120 \\
64  & 12577 & 5 & 0.0370 $\pm$ 0.0096 & 0.0111 $\pm$ 0.0026 & 0.0553 $\pm$ 0.0122 \\
128 & 25057 & 5 & 0.0398 $\pm$ 0.0057 & 0.0117 $\pm$ 0.0017 & 0.0628 $\pm$ 0.0102 \\
\hline
\end{tabular}
\end{center}
\end{table}

The ablation shows that the 128-neuron model is not uniquely responsible for the reconstruction signal: all four hidden sizes give comparable midpoint and duration errors, and the 64-neuron model gives the lowest mean MAE in this auxiliary test. Nevertheless, because all runs use the same small simulation suite, the ablation should be interpreted only as an internal robustness check rather than as evidence of out-of-prior generalization.

\section{RESULTS}

In this section we present the reconstructed reionization history $x_{\mathrm{HI}}(z)$ inferred from fixed-$k$ trajectories of \(\Delta^2_{21}(k,z)\). Unless stated otherwise, the input trajectories are taken directly from the simulated signal, so the discussion first isolates the structure of the inverse map before observational perturbations are added.

\subsection{Recovered reionization history}

\begin{figure}[!htbp]
\centering
\begin{subfigure}{0.48\textwidth}
\centering
\includegraphics[width=\textwidth]{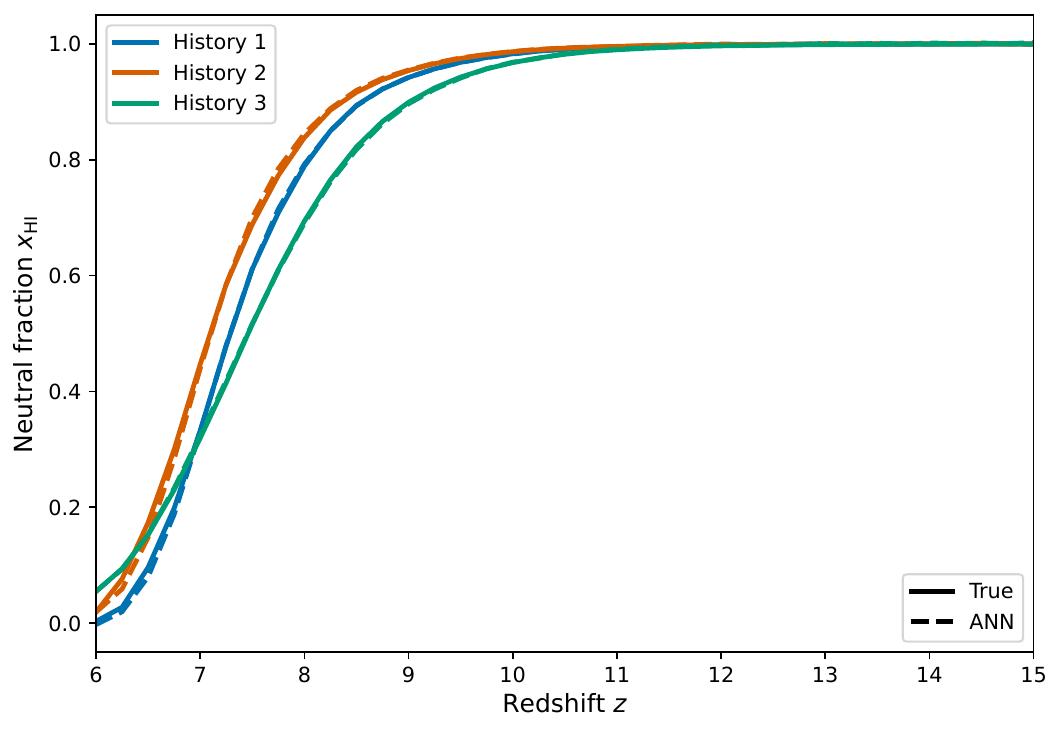}
\caption{$k = 0.1~h\,\mathrm{Mpc}^{-1}$}
\end{subfigure}
\hfill
\begin{subfigure}{0.48\textwidth}
\centering
\includegraphics[width=\textwidth]{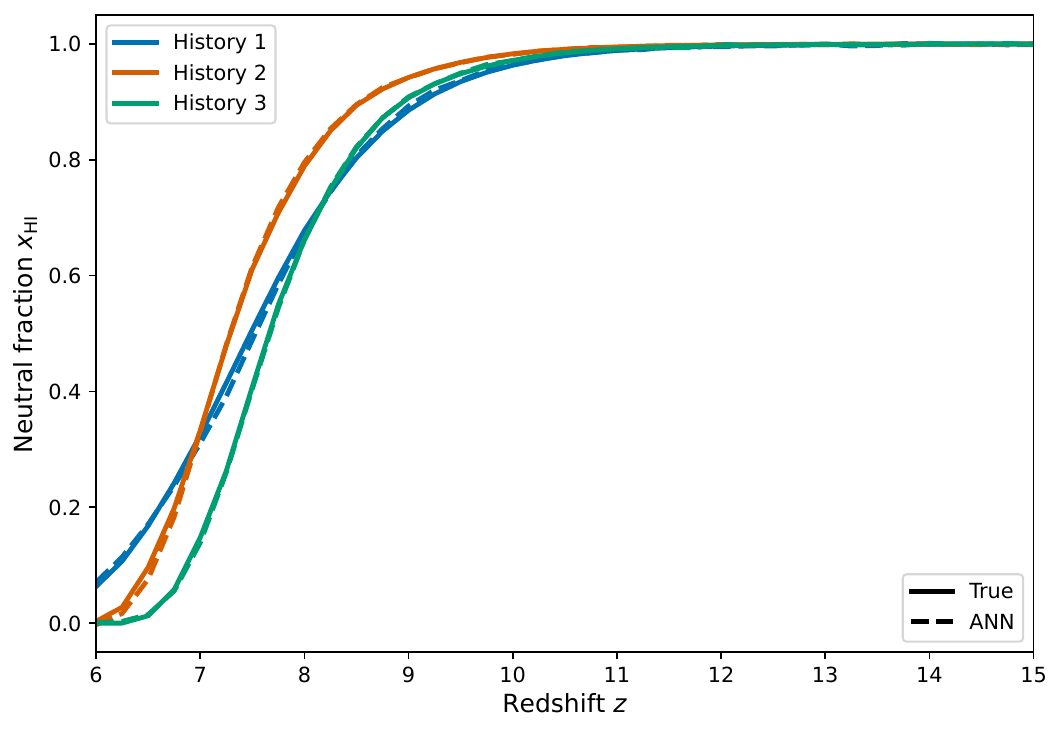}
\caption{$k = 0.5~h\,\mathrm{Mpc}^{-1}$}
\end{subfigure}

\vspace{0.4cm}

\begin{subfigure}{0.68\textwidth}
\centering
\includegraphics[width=\textwidth]{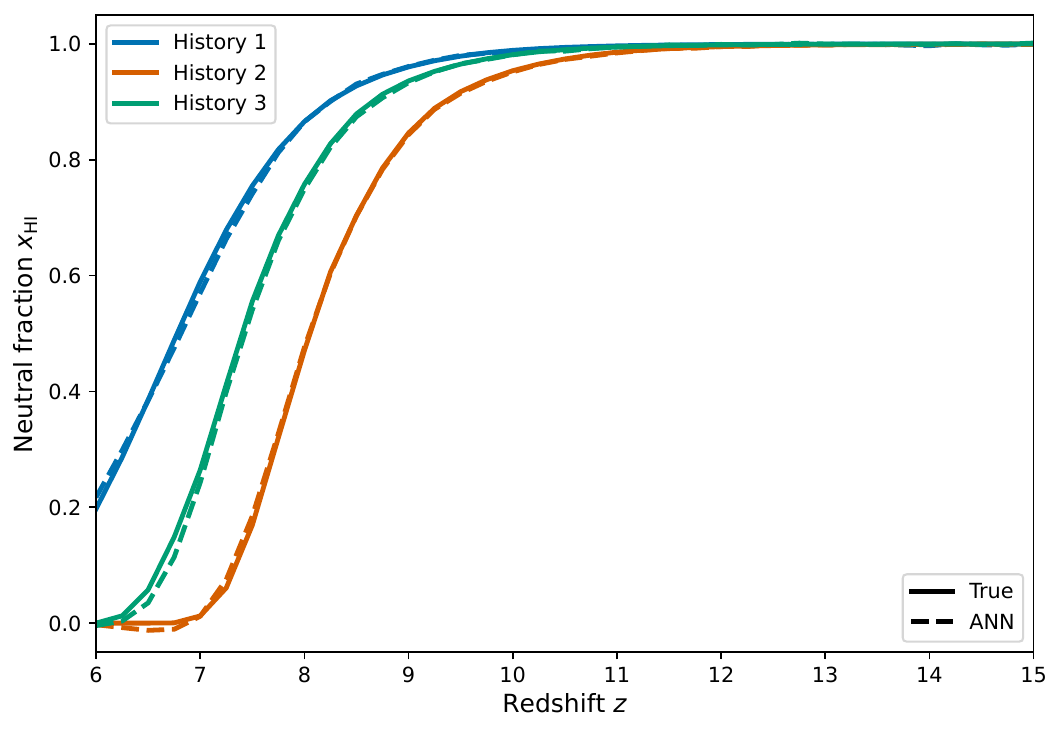}
\caption{$k = 1.0~h\,\mathrm{Mpc}^{-1}$}
\end{subfigure}
\caption{Reconstructed reionization histories using fixed-$k$ trajectories of \(\Delta^2_{21}(k,z)\) at three representative scales. Each panel overlays three illustrative test histories; colors denote different histories, while solid and dashed curves denote the true and ANN-reconstructed histories, respectively.}
\label{fig:5}
\end{figure}

The ANN reconstructs $x_{\rm HI}(z)$ on the full 97-bin grid over $z=6$--$15$
using the dimensionless power spectrum evaluated at fixed wavenumber.
The figures below display this full EoR-relevant redshift range. To assess the scale dependence of the inverse mapping, we compare reconstructions obtained at $k=0.1$, $0.5$, and $1.0~h\,\mathrm{Mpc}^{-1}$. Figure~\ref{fig:5} presents three representative reconstructed histories for each of these three cases, with colors distinguishing different test histories and line styles distinguishing the true and ANN-reconstructed curves. One generally expects the recovery to become less accurate toward larger $k$, because smaller-scale power is increasingly sensitive to local ionization morphology and is therefore less directly coupled to the global progress of reionization.

Within the present dataset, however, the reconstructions remain strong even at $k = 1.0~h\,\mathrm{Mpc}^{-1}$. We do not interpret this as evidence that small-scale modes are generically sufficient. Rather, it reflects the restricted training family adopted here: once the astrophysical parameter space is compressed to three parameters, the allowed histories occupy a relatively narrow manifold of smooth curves, and even higher-$k$ power-spectrum trajectories retain useful timing information. The $k=1.0~h\,\mathrm{Mpc}^{-1}$ case should therefore be understood as a simulation-based information-content test of the inverse mapping, not as a claim that the fiducial SKA1-Low-like core configuration adopted in the noise section can measure that scale with high sensitivity. The apparently weak scale dependence should therefore be understood as a prior-dependent feature of the present setup.

\subsection{Correlation-based metric}

To summarize the similarity between the reconstructed and true histories on the independent test set, we use the correlation coefficient (CC),
\begin{equation}
\label{eq:cc}
CC = \frac{\sum_{i=1}^{N_z} (y_{\text{true},i} - \bar{y}_{\text{true}})(y_{\text{ANN},i} - \bar{y}_{\text{ANN}})}{\sqrt{\sum_{i=1}^{N_z} (y_{\text{true},i} - \bar{y}_{\text{true}})^2} \sqrt{\sum_{i=1}^{N_z} (y_{\text{ANN},i} -\bar{y}_{\text{ANN}})^2}},
\end{equation}
where $y_{\text{true},i}$ and $y_{\text{ANN},i}$ denote the true and reconstructed neutral fractions at redshift $z_i$, and $N_z$ is the number of redshift bins. The overbars denote redshift averages. Although the Pearson correlation coefficient is often introduced in the context of linear relations and linear fitting, its normalized form can also be used as a profile-reproduction diagnostic for two discretized histories. We use it in this latter sense: after the true and reconstructed histories are sampled on the same redshift grid, CC measures how well their mean-subtracted variations are aligned across redshift. A high CC therefore indicates that the ANN reproduces the overall redshift-dependent shape of the target history, but it should not be interpreted as implying that $x_{\mathrm{HI}}(z)$ itself is linear in redshift.

This use is justified only as a complementary measure of reconstruction fidelity, because CC is insensitive to some errors that are physically important for reionization. In particular, smooth monotonic histories can yield high CC values even when the midpoint or duration of the transition is imperfectly recovered, and an affine rescaling of a curve can leave CC nearly unchanged. We therefore use CC to summarize global shape agreement, while MAE, RMSE, $z_{50}$, and $\Delta z$ are evaluated separately below to quantify absolute reconstruction accuracy and transition-specific errors.

\begin{figure}[!htbp]
\centering
\begin{subfigure}{0.48\textwidth}
\centering
\includegraphics[width=\textwidth]{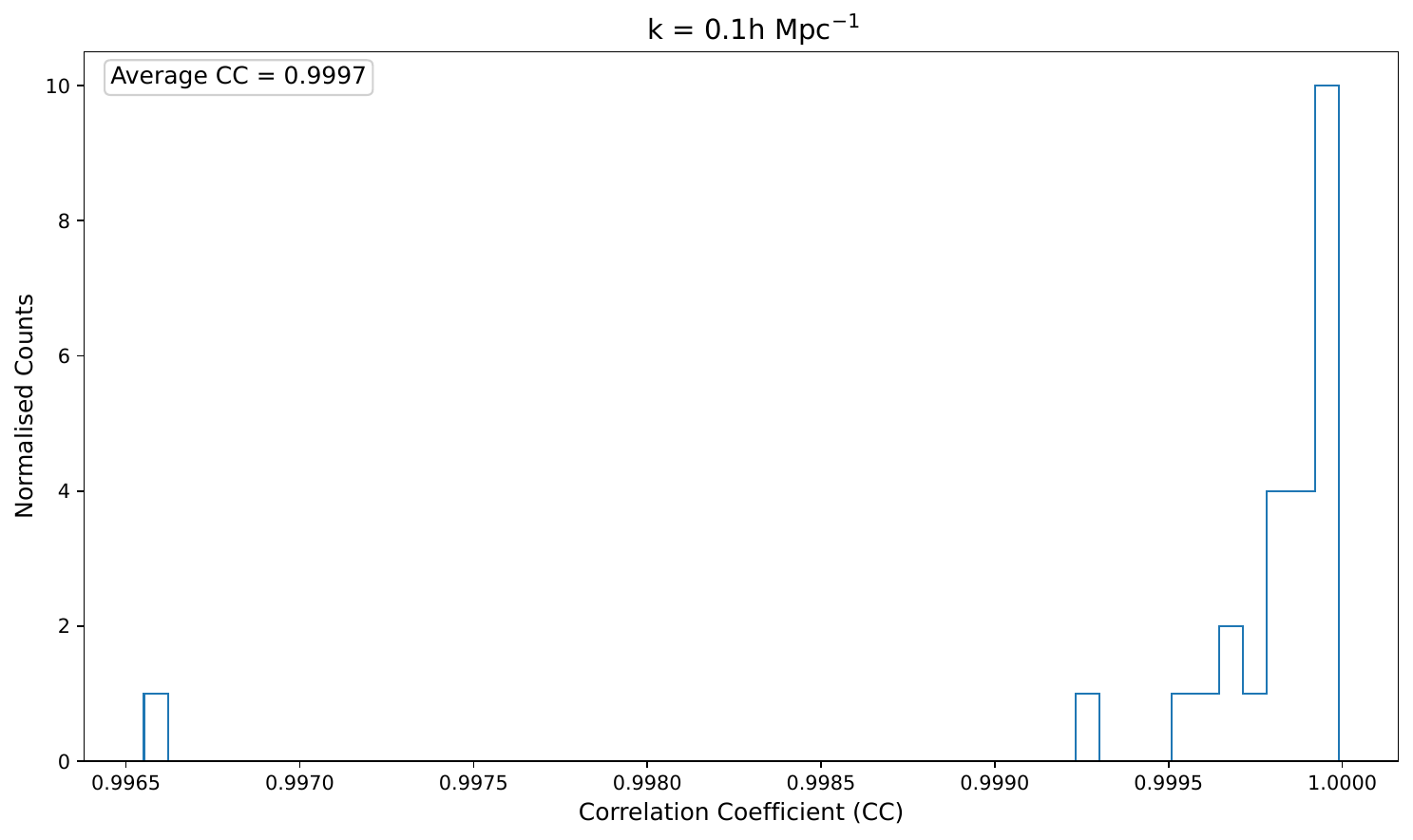}
\caption{$k = 0.1~h\,\mathrm{Mpc}^{-1}$}
\end{subfigure}
\hfill
\begin{subfigure}{0.48\textwidth}
\centering
\includegraphics[width=\textwidth]{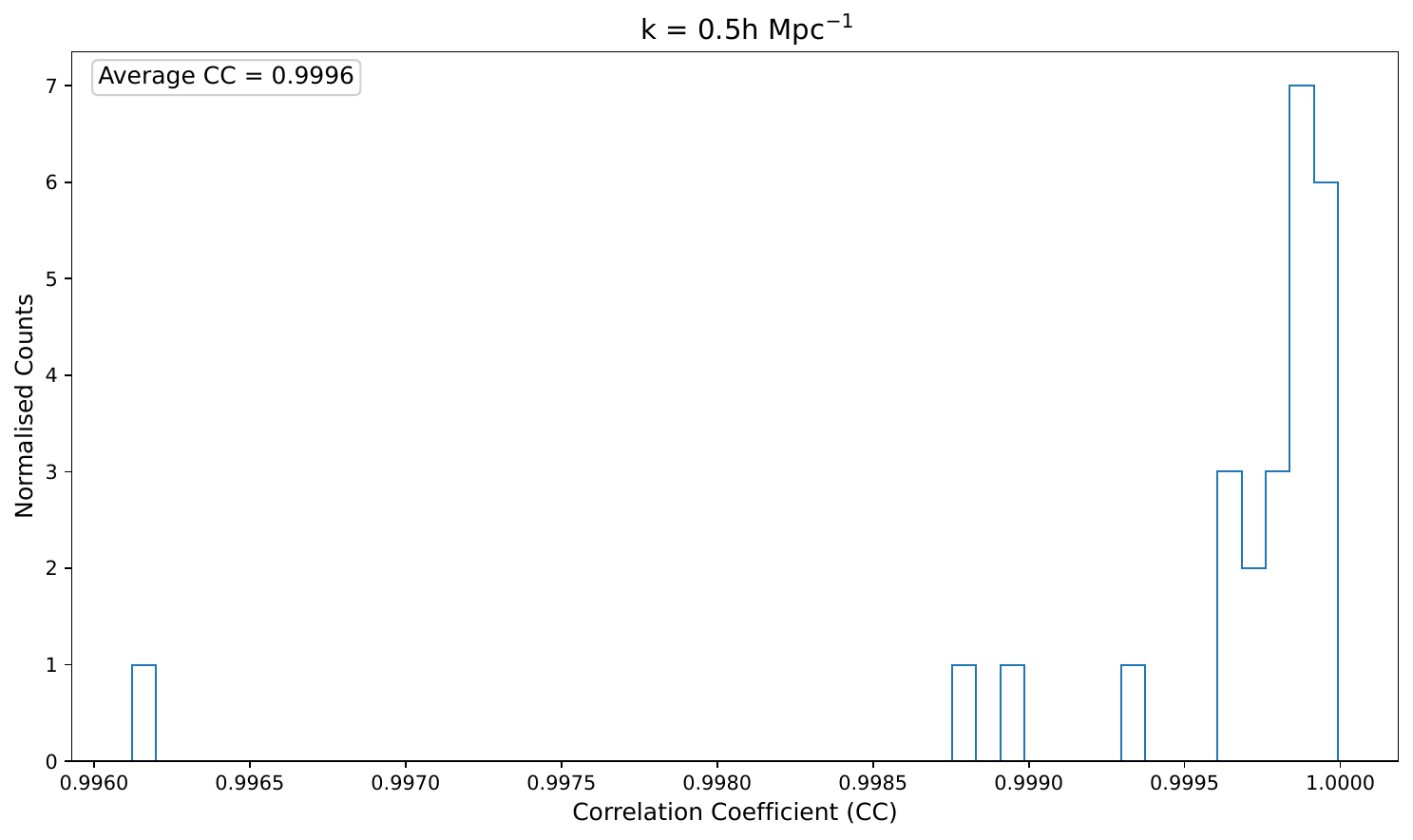}
\caption{$k = 0.5~h\,\mathrm{Mpc}^{-1}$}
\end{subfigure}

\vspace{0.4cm}

\begin{subfigure}{0.68\textwidth}
\centering
\includegraphics[width=\textwidth]{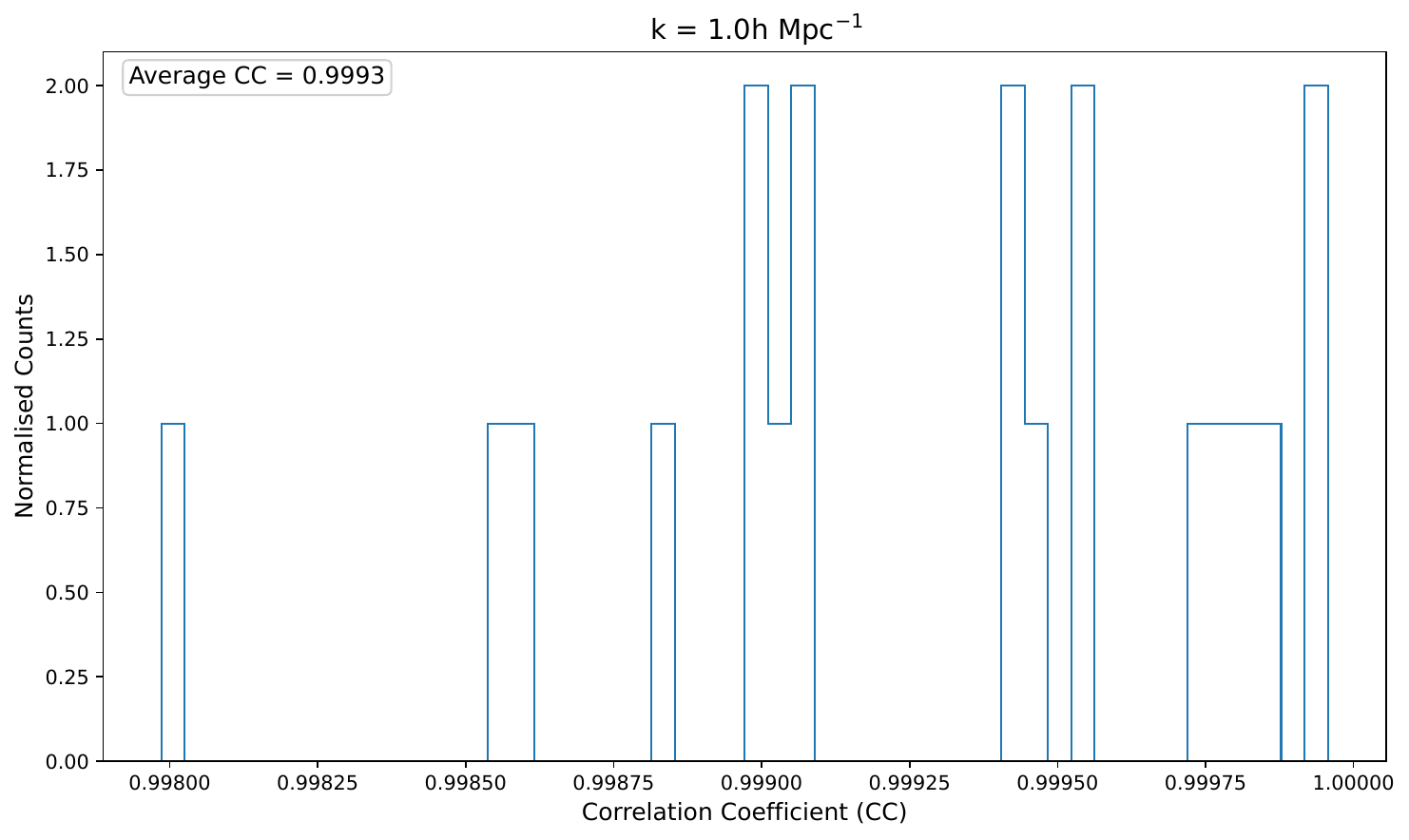}
\caption{$k = 1.0~h\,\mathrm{Mpc}^{-1}$}
\end{subfigure}
\caption{Distribution of correlation coefficients for reconstructions from fixed-$k$ \(\Delta^2_{21}(k,z)\) trajectories at three representative scales.}
\label{fig:7}
\end{figure}

Using CC in this complementary sense, we next summarize how the global shape agreement changes with scale. Figure~\ref{fig:7} shows the CC distributions for the reconstructed reionization histories obtained from the redshift evolution of the fixed-$k$ power spectrum at $k = 0.1$, $0.5$, and $1.0~h\,\mathrm{Mpc}^{-1}$. For the two lower-$k$ cases, corresponding to larger spatial scales, the distributions are narrowly concentrated at high CC values, consistent with the visual impression in Figure~\ref{fig:6} and with the expectation that these modes are more directly linked to the growth and topology of ionized regions.

At $k = 1.0~h\,\mathrm{Mpc}^{-1}$, the CC values also remain high, but this should not be interpreted as evidence for reconstruction quality comparable to the lower-$k$ cases. As already seen in Figure~\ref{fig:6}, the recovered histories at this scale show visibly larger scatter and less faithful tracking of the transition region. This apparent tension reflects the limitation discussed above: CC mainly measures agreement in the overall monotonic profile and is relatively insensitive to moderate mismatches in the exact midpoint and width of the transition. Within the restricted three-parameter prior adopted here, most histories are smooth and share a similar global form, so even noticeably degraded reconstructions can still yield high CC values. We therefore interpret the $k=1.0~h\,\mathrm{Mpc}^{-1}$ result as showing that the network can still recover the broad reionization trend within our restricted three-parameter simulation set, rather than as evidence that high-$k$ power spectra alone generally provide a precise reconstruction of the reionization history.

\subsection{Recovery with idealized observational noise}

So far we have considered the noise-free case, in which the input power-spectrum trajectories contain no observational noise. However, real observations are affected by thermal noise, foreground contamination, calibration systematics, and other instrumental effects. In this section, we consider a more realistic but still idealized case by adding noise to the 21\,cm power spectrum before using it as the ANN input. Specifically, we include only a foreground-free combination of thermal noise and sample variance in order to test how stable the inverse mapping remains under a limited observational perturbation model.

The sensitivity of a radio interferometer to the 21cm power spectrum is determined by the array configuration, integration time, and system temperature. Following the standard formalism derived in previous studies \citep{2006ApJ...653..815M,2008PhRvD..78b3529M,2012ApJ...756..165P,2013PhRvD..88h1303M}, we express the thermal noise power spectrum for a single Fourier mode $\mathbf{k}$ as:

\begin{equation}
    P_{\text{th,1mode}}(k, \mu) \approx \frac{X^2 Y \Omega_{\text{fov}} T_{\text{sys}}^2}{t_{\text{int}} n(k_\perp) A_e},
    \label{eq:pth}
\end{equation}

\noindent where $X^2 Y$ converts the interferometric bandwidth and solid angle into comoving volume units (see \citealt{2006PhR...433..181F}), with $X \approx d_A^2$ (comoving angular diameter distance squared) and $Y \approx \lambda_{21}(1+z)^2/H(z)$. $\Omega_{\text{fov}} \approx \lambda^2 / A_e$ is the solid angle of the primary beam field of view, $A_e$ is the effective collecting area per station, and $t_{\text{int}}$ represents the total integration time. $n(k_\perp)$ denotes the number density of baselines in the $uv$ plane that are sensitive to the transverse wavenumber $k_\perp = k\sqrt{1-\mu^2}$, where $\mu = \cos\theta$ is the cosine of the angle between the wavevector $\mathbf{k}$ and the line of sight. A high baseline density in the array core significantly enhances sensitivity to large scale structures (e.g., \citealt{2012ApJ...756..165P, 2015aska.confE...1K}).

In this expression, $\Omega_{\text{fov}}$ should be understood as the instantaneous primary-beam field of view for a single pointing, corresponding to $N_b=1$ in the notation of \citet{2015aska.confE...1K}. As discussed by those authors, SKA1-Low observing strategies can make use of multi-beaming; for example, the bandwidth can be split into two beams, and increasing the effective field of view through multi-beaming or drift-scanning can reduce the sample-variance contribution on large angular scales. More generally, if an observing program covers several independent fields---for example the four-field case mentioned by the referee---the effective survey volume entering the mode count increases, and the sample-variance term would decrease approximately as $N_{\rm field}^{-1/2}$ for independent fields of comparable depth. Our fiducial noise realization deliberately keeps $N_b=1$ and therefore uses a single-field/effective one-pointing baseline. A multi-field or multi-beam survey would be more favorable with respect to survey volume and sample variance, but modelling such strategies requires specifying the number of fields or beams and the allocation of integration time per field. Since the purpose here is a reproducible proof-of-concept perturbation test rather than an optimized SKA survey forecast, we leave multi-field/multi-beam survey optimization to future work.

The system temperature, $T_{\text{sys}}$, is the sum of the receiver temperature and the sky brightness temperature:
\begin{equation}
    T_{\text{sys}}(\nu) = T_{\text{rcvr}} + T_{\text{sky}}(\nu).
\end{equation}
At the low frequencies relevant to the Epoch of Reionization (EoR) and Cosmic Dawn, the system noise is dominated by the Galactic synchrotron background. We model the sky temperature as a power law following \cite{1982A&AS...47....1H} and \cite{2017PASP..129d5001D}:
\begin{equation}
    T_{\text{sky}}(\nu) \approx 60 \left(\frac{\nu}{300\,\text{MHz}}\right)^{-2.55} \,\text{K}.
\end{equation}
We assume a receiver temperature of $T_{\text{rcvr}} \approx 100$\,K, consistent with SKA1-Low design specifications \citep{5136190, 2019arXiv191212699B}.

Since the 21cm signal is isotropic (after accounting for redshift space distortions), we estimate the spherically averaged power spectrum sensitivity in a spherical shell of width $\Delta k$. The variance of the thermal noise for a bin centered at $k$ is given by the inverse variance weighting of all modes in the shell \citep{Lidz_2011, 2013PhRvD..88h1303M}:
\begin{equation}
    P_{\text{thermal}}(k) = \left[ \sum_{\mu} \frac{N_c(k, \mu)}{P_{\text{th,1mode}}^2(k, \mu)} \right]^{-1/2},
\end{equation}
where $N_c(k, \mu)$ is the number of independent modes in the $k$ shell region defined by $k$, $\mu$, and logarithmic bin width $\epsilon = \Delta k / k$. Specifically, $N_c \approx \epsilon k^3 \Delta \mu \cdot V_{\text{survey}} / (4\pi^2)$, where $V_{\text{survey}}$ is the survey volume.

In addition to thermal noise, the precision of power spectrum measurements is fundamentally limited by cosmic variance (sample variance), which arises from the finite volume surveyed. This is particularly dominant at large scales (small $k$). The error contribution from cosmic variance is estimated as:
\begin{equation}
    P_{\text{cv}}(k) = \frac{P_{21}(k)}{\sqrt{N_{\text{modes}}}},
\end{equation}
where $N_{\text{modes}}$ is the total number of independent Fourier modes in the upper half of the $k$ shell.

The total observational uncertainty is the quadrature sum of the thermal noise and cosmic variance:
\begin{equation}
    \sigma_P(k) = \sqrt{P_{\text{thermal}}^2(k) + P_{\text{cv}}^2(k)}.
\end{equation}
Although the sensitivity expressions above are conventionally written for $P_{21}$, the actual ANN inputs are the corresponding dimensionless trajectories $\Delta^2_{21}(k,z)$. We therefore construct the noisy inputs in that same normalization by adding a Gaussian realization to the dimensionless signal trajectory:
\begin{equation}
    \Delta^2_{\mathrm{obs}}(k,z) = \Delta^2_{21}(k,z) + N_{\Delta^2}(k,z).
\end{equation}

For the noise realization, we adopt parameters representative of an idealized foreground-free SKA1-Low-like sensitivity calculation targeting the EoR \citep{5136190, 2019arXiv191212699B,2025RAA....25h5017S}. The adopted observational specifications and analysis assumptions are summarized in Table~\ref{tab:ska_noise_spec}.

\begin{table}[!htbp]
\begin{center}
\caption[]{SKA1-Low-like observational specifications adopted for the noise model.}\label{tab:ska_noise_spec}
\small
\begin{tabular}{@{}p{3.1cm}p{9.0cm}@{}}
\hline\noalign{\smallskip}
Item & Specification \\
\hline\noalign{\smallskip}
Array layout & Compact 224-station hexagonal core within radius $500$\,m (core diameter $\sim 1000$\,m), with minimum baseline $60$\,m and maximum retained baseline $1000$\,m. This compact-core setup is intended for the illustrative SKA-like noise realization at large EoR scales and should not be read as guaranteeing sensitivity to all simulated modes, especially $k=1.0~h\,\mathrm{Mpc}^{-1}$. \\
Primary beam & Gaussian beam with reference frequency $150$\,MHz and dish size $35$\,m. \\
Site and receiver & Observatory latitude $-26.8^\circ$ and receiver temperature $T_{\mathrm{rcvr}}=100$\,K. \\
Integration time & Total $t_{\mathrm{int}} = 1000$ hours, implemented as 100 observing days with 10 hours per day. \\
Field of view & Single-beam FWHM $\sim 3.5^\circ$ at $z\sim 8$. This fiducial noise realization uses a single-beam/single-field assumption ($N_b=1$, $N_{\rm field}=1$); the additional survey-volume gain from multi-field or SKA multi-beaming strategies is not included. \\
Bandwidth & $B = 10$\,MHz per redshift bin. \\
Effective area & $A_e \approx 421\,\mathrm{m}^2$ at $z \sim 8$, scaling with frequency as $\lambda^2$. \\
$k$-binning & Logarithmic step size $\epsilon = \delta k / k = 0.1$. \\
Noise assumption & Foreground-free sensitivity calculation, with random Gaussian realizations drawn from the 1D thermal-plus-sample-variance uncertainty at the target $k$. \\
\noalign{\smallskip}\hline
\end{tabular}
\end{center}
\end{table}

For each redshift, we first calculate the sensitivity at $\nu=1420/(1+z)$ MHz, extract the nearest 1D sensitivity bin to the target $k$, and draw one Gaussian realization for the SKA-like case. Our noise estimates are consistent with previous studies (e.g., \citealt{2015aska.confE...1K}). We emphasize that this SKA-like noise realization is used only as an illustrative perturbation model for the representative large-scale reconstruction, and the compact-core configuration in Table~\ref{tab:ska_noise_spec} is not intended to demonstrate practical detectability at every simulated wavenumber. In particular, the $k=1.0~h\,\mathrm{Mpc}^{-1}$ reconstruction shown above remains part of the noise-free scale-dependence test, not a forecast for the fiducial $\sim1000$ m core. For the idealized SKA1-Low-like setup adopted here, the signal remains larger than the thermal-noise contribution across much of the redshift range for the representative large-scale case considered. Within this favorable signal-to-noise regime, and still within the compressed three-parameter model family, the inverse mapping from $\Delta^2_{21}(k,z)$ to $x_{\mathrm{HI}}(z)$ remains reasonably stable.

We also examine reconstructed histories over the displayed range $z=6$--$15$ when noise is added according to the idealized SKA1-Low-like specification, together with an amplified case in which the thermal-noise amplitude is multiplied by $10^3$ as a stress test of a much less favorable regime. In other words, the ``$1000\times$ SKA'' case denotes the same SKA-like noise model but with the noise standard deviation scaled up by a factor of $10^3$. This amplified case is not intended as a quantitative model of any specific instrument.

\begin{figure}[!htbp]
\centering
\includegraphics[width=14cm]{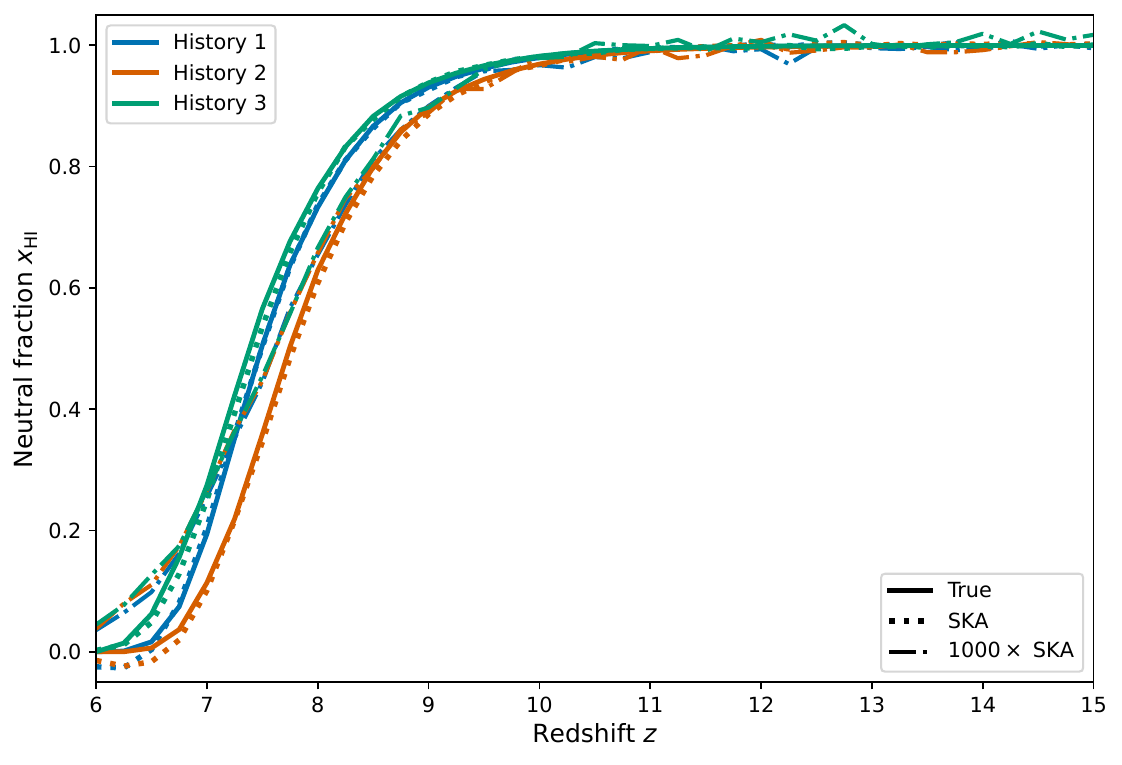}
\caption{Illustrative reconstructions of the neutral-fraction history from noisy fixed-$k$ power-spectrum input over $z=6$--$15$. Colors denote three different test histories. For each history, the solid curve shows the true neutral-fraction history, the dotted curve shows the reconstruction obtained with the idealized SKA1-Low-like noise realization, and the dash-dotted curve shows the reconstruction for the exaggerated ``$1000\times$ SKA'' stress test, i.e., the same SKA-like noise model with the noise amplitude increased by a factor of $10^3$.}
\label{fig:9}
\end{figure}

Figure~\ref{fig:9} shows three illustrative reconstructions of $x_{\mathrm{HI}}(z)$ from noisy fixed-$k$ power-spectrum input. For the idealized SKA-like noise level, the reconstructed histories remain close to the true curves throughout the transition in these examples, indicating that the network can recover the overall monotonic form and an approximately correct midpoint in a favorable regime. By contrast, the $1000\times$ noise stress test shows visibly larger deviations, although it still preserves the coarse S-shaped trend. This visual comparison is intended only as an illustration; the more systematic error analysis below is based on the full held-out test subset.

Because the restricted parameter space produces smooth S-shaped histories, whole curve metrics alone may overstate what the network has learned. The physically most informative regime is the transition interval, where the ionized fraction changes rapidly. We therefore re-evaluate the reconstruction specifically in the range $0.2 \leq x_{\mathrm{HI}} \leq 0.8$.

We evaluate three physically meaningful aspects of the reconstruction within this interval: the neutral-fraction history $x_{\mathrm{HI}}(z)$ itself, the reionization midpoint $z_{50}$, and the duration $\Delta z$. For the history, we quantify the reconstruction error using the mean absolute error (MAE) and the root-mean-square error (RMSE) computed from the $x_{\mathrm{HI}}$ residuals within the transition interval. We also track the signed residual mean (bias), median, and standard deviation. For the derived summary quantities, we evaluate dimensionless relative errors, i.e. the midpoint and duration residuals are normalized by the corresponding true values.

MAE measures the mean absolute difference between the predicted and true neutral fractions:
\begin{equation}
\text{MAE} = \frac{1}{N_{\text{test}}N_z} \sum_{n=1}^{N_{\text{test}}} \sum_{i=1}^{N_z} \left| y_{\text{true}, i}^{(n)} - y_{\text{ANN}, i}^{(n)} \right|.
\end{equation}

RMSE is defined as
\begin{equation}
\text{RMSE} = \sqrt{ \frac{1}{N_{\text{test}}N_z} \sum_{n=1}^{N_{\text{test}}} \sum_{i=1}^{N_z} \left( y_{\text{true}, i}^{(n)} - y_{\text{ANN}, i}^{(n)} \right)^2 }.
\end{equation}
Like MSE, RMSE gives greater weight to large residuals than MAE.

For the neutral-fraction history, we define the signed residual as
$y_{\text{true}}-y_{\text{ANN}}$, so positive values indicate that the ANN
underestimates $x_{\mathrm{HI}}$ on average. For the derived summary
quantities, we instead define signed errors as predicted minus true.

The midpoint error is defined as the relative error
\begin{equation}
\label{eq:errz50}
\mathrm{Error}_{z_{50}}
= \frac{z_{50}^{\mathrm{ANN}} - z_{50}^{\mathrm{true}}}{z_{50}^{\mathrm{true}}} .
\end{equation}
For the duration, we use the same positive-duration convention as in
Equation~\ref{eq:deltazdef},
\begin{equation}
\Delta z \equiv z_{75}-z_{25},
\end{equation}
where $x_{\mathrm{HI}}(z_{75})=0.75$ and
$x_{\mathrm{HI}}(z_{25})=0.25$. The duration error is then defined as the relative error
\begin{equation}
\label{eq:errdz}
\mathrm{Error}_{\Delta z}
= \frac{\Delta z^{\mathrm{ANN}} - \Delta z^{\mathrm{true}}}{\Delta z^{\mathrm{true}}} .
\end{equation}

\begin{figure}[!htbp]
\centering
\includegraphics[width=14cm]{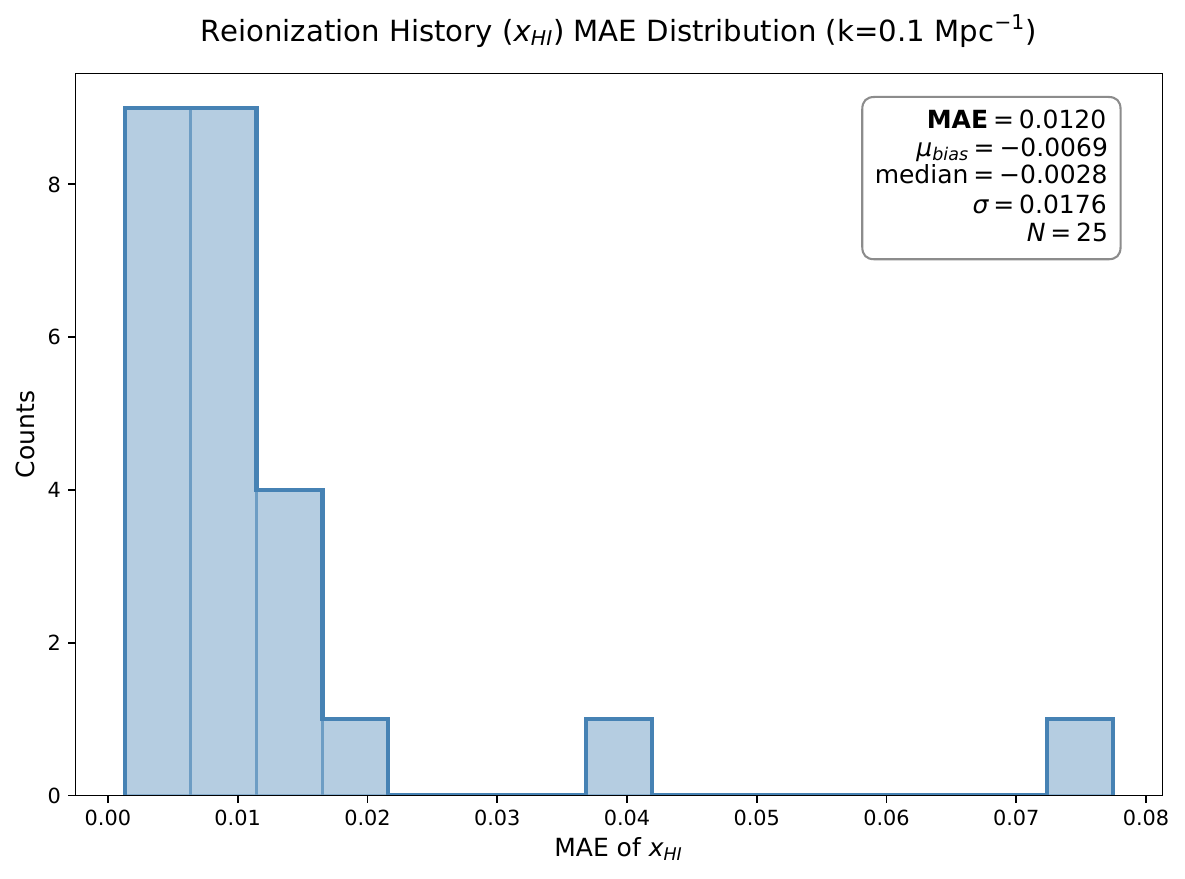}
\caption{MAE distribution for the reconstructed reionization history in the transition interval using noisy inputs at $k = 0.1~h\,\mathrm{Mpc}^{-1}$.}
\label{fig:10}
\end{figure}

\begin{figure}[!htbp]
\centering
\includegraphics[width=14cm]{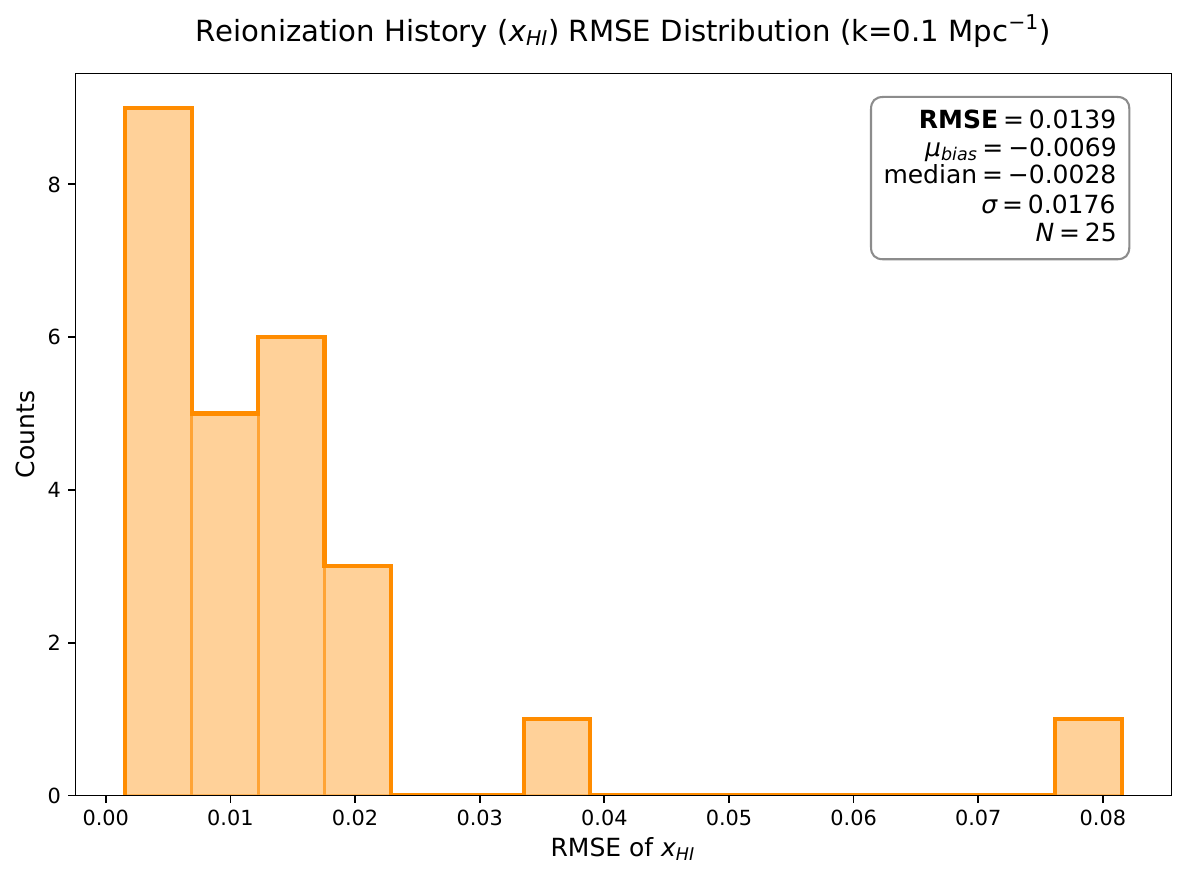}
\caption{RMSE distribution for the reconstructed reionization history in the transition interval using noisy inputs at $k = 0.1~h\,\mathrm{Mpc}^{-1}$.}
\label{fig:11}
\end{figure}

\begin{figure}[!htbp]
\centering
\includegraphics[width=14cm]{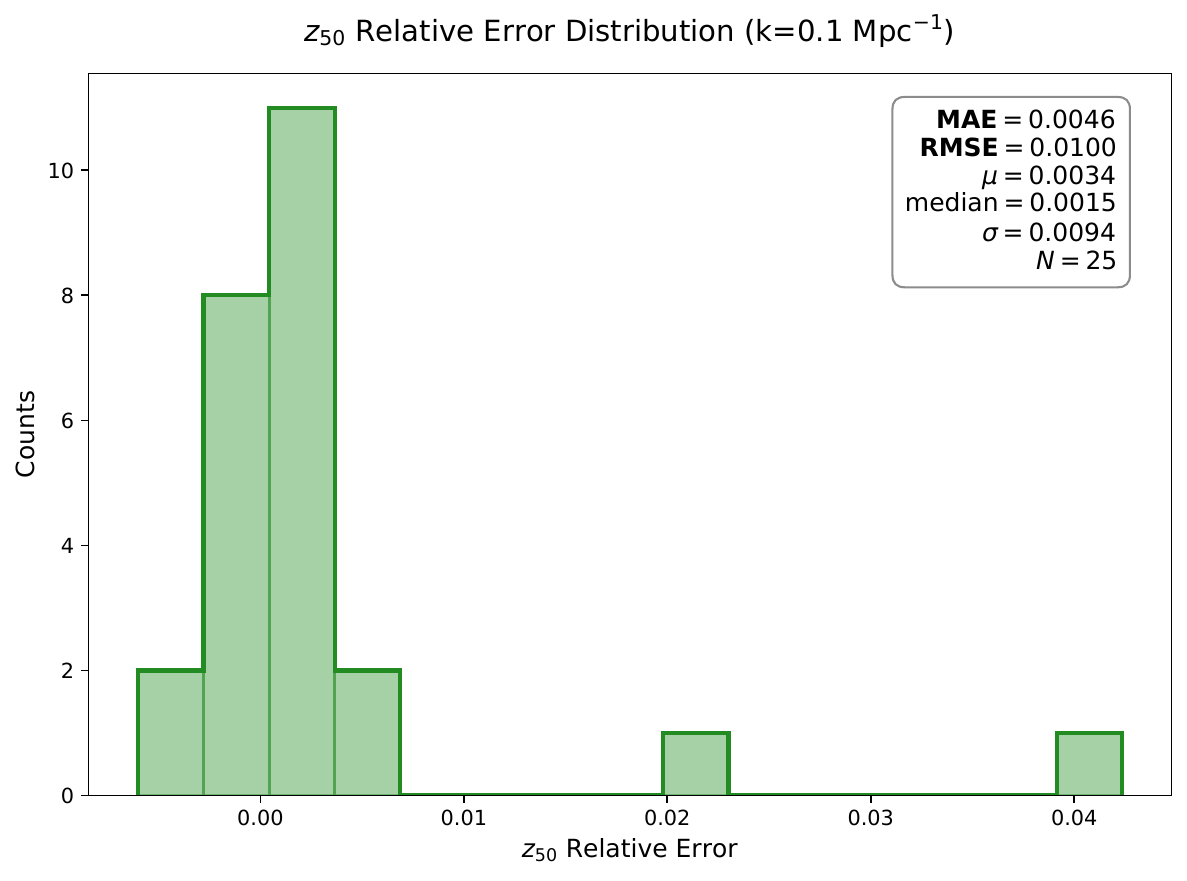}
\caption{Distribution of the signed relative midpoint error defined by Equation~\ref{eq:errz50}.}
\label{fig:12}
\end{figure}

\begin{figure}[!htbp]
\centering
\includegraphics[width=14cm]{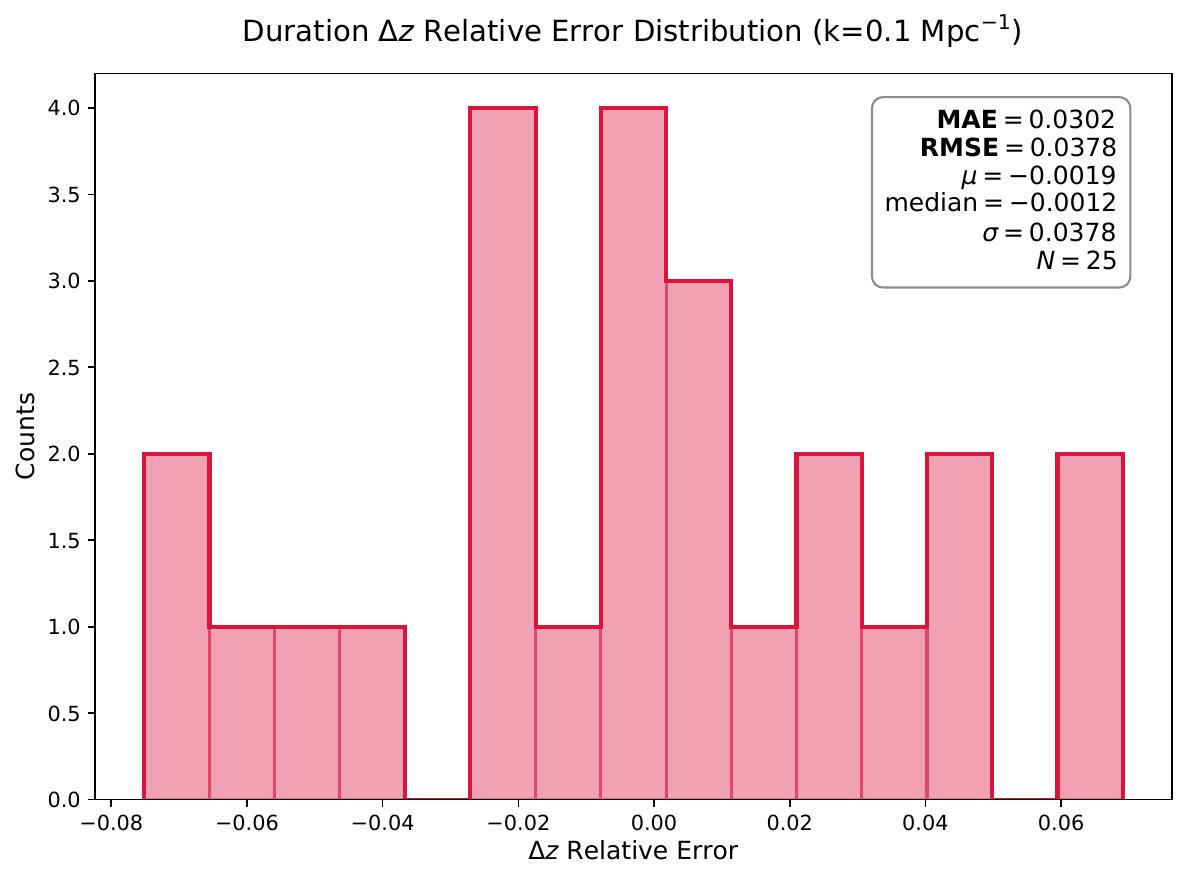}
\caption{Distribution of the signed relative duration error defined by Equation~\ref{eq:errdz}.}
\label{fig:13}
\end{figure}

\begin{figure}[!htbp]
\centering
\includegraphics[width=14cm]{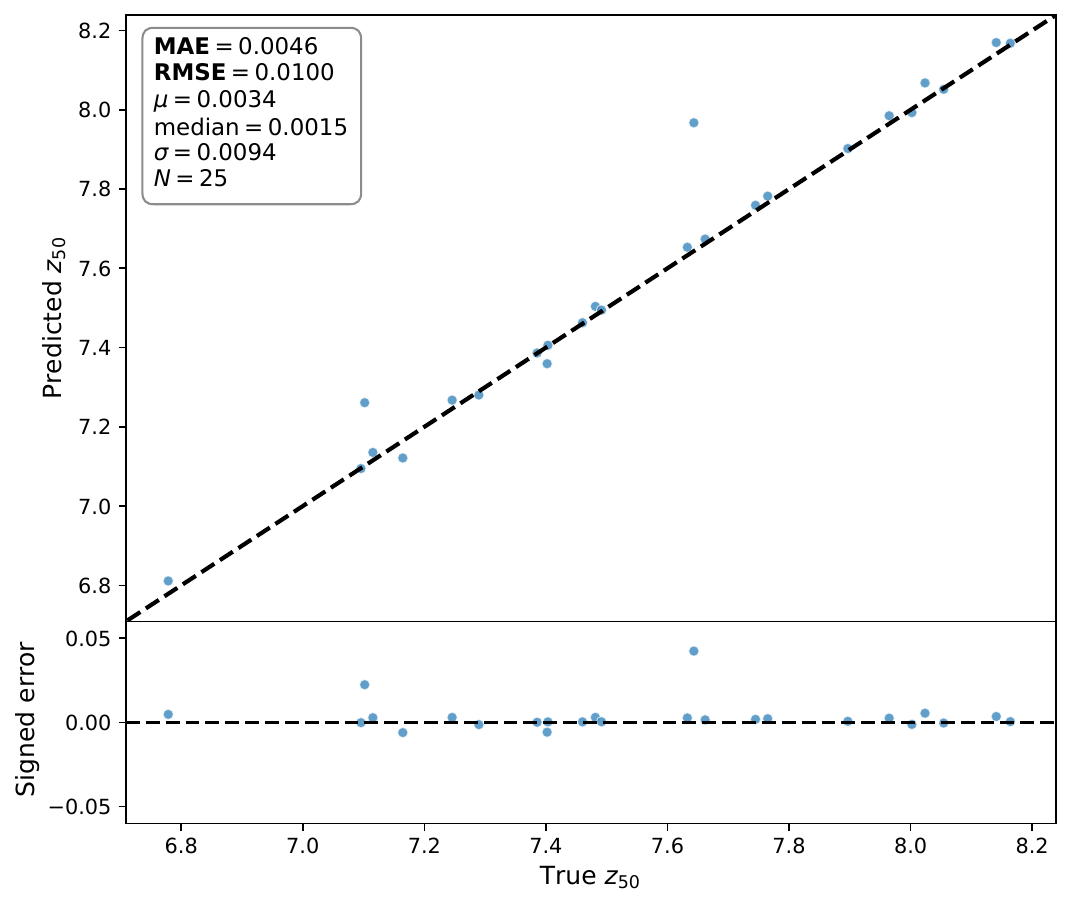}
\caption{Scatter plot of midpoint recovery for the reconstructed reionization history. The lower panel shows the signed relative midpoint error defined by Equation~\ref{eq:errz50}.} 
\label{fig:14}
\end{figure}

\begin{figure}[!htbp]
\centering
\includegraphics[width=14cm]{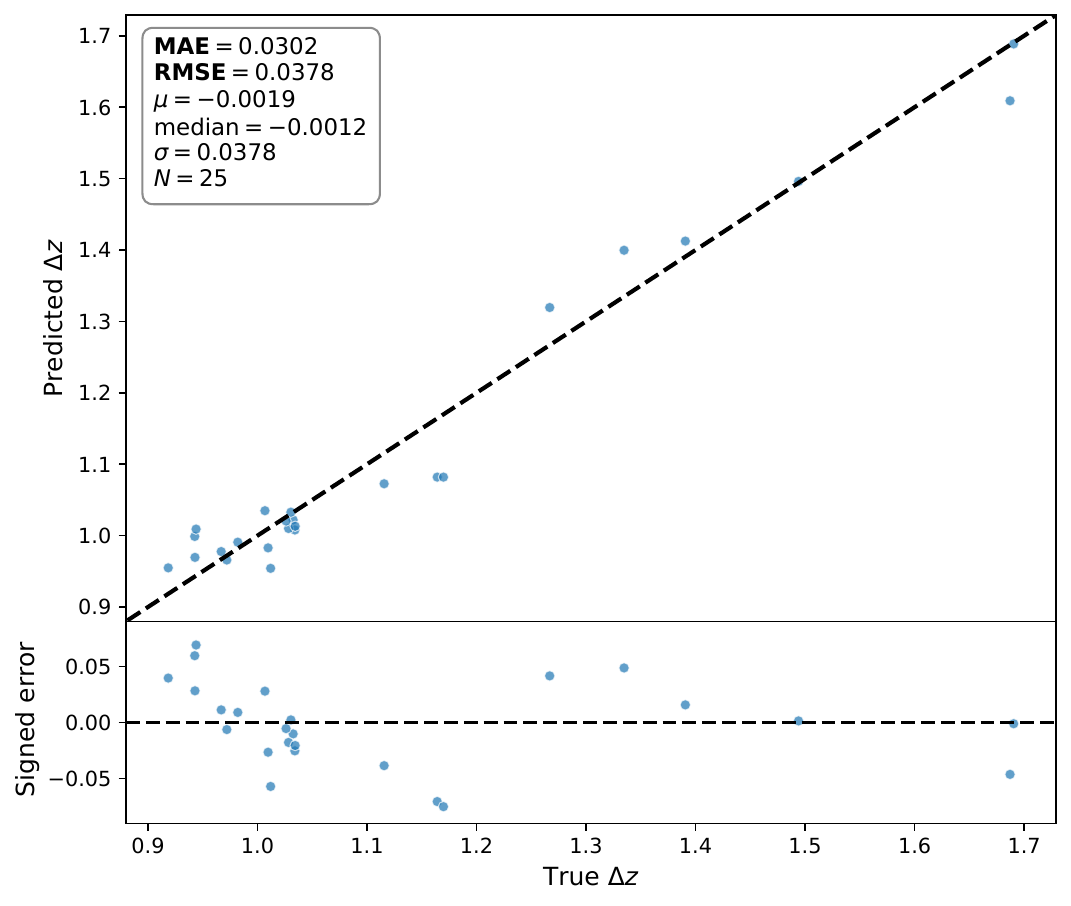}
\caption{Scatter plot of duration recovery for the reconstructed reionization history. The lower panel shows the signed relative duration error defined by Equation~\ref{eq:errdz}.}
\label{fig:15}
\end{figure}

Figures~\ref{fig:10}--\ref{fig:15} summarize the error distributions in the transition regime, and Table~\ref{tab:performance} collects the main statistics.

%
\begin{table}
\begin{center}
\caption[]{Representative performance summary for one independent test subset ($N=25$ trajectories) for the three main reionization descriptors. For $x_{\rm HI}$, the listed statistics are absolute neutral-fraction errors. For $z_{50}$ and $\Delta z$, they are dimensionless relative errors defined by \mbox{Equations~\ref{eq:errz50} and~\ref{eq:errdz}}.\label{tab:performance}}


 \begin{tabular}{lcccccc}
  \hline\noalign{\smallskip}
Parameter & MAE & RMSE & Bias ($\mu$) & Median & $\sigma$ & $N$ \\
  \hline\noalign{\smallskip}
Reionization History ($x_{\rm HI}$) & 0.0120 & 0.0139 & -0.0069 & -0.0028 & 0.0176 & 25 \\
Reionization Midpoint ($z_{50}$)    & 0.0046 & 0.0100 & 0.0034 & 0.0015 & 0.0094 & 25 \\
Reionization Duration ($\Delta z$)               & 0.0302 & 0.0378 & -0.0019 & -0.0012 & 0.0378 & 25 \\
  \noalign{\smallskip}\hline
\end{tabular}
\end{center}
\end{table}

Table~\ref{tab:performance} summarizes representative one-split performance for the three key reionization descriptors,
evaluated within the physically most informative transition interval $0.2 \leq x_{\mathrm{HI}} \leq 0.8$ for the representative noisy $k = 0.1\,h\,\mathrm{Mpc}^{-1}$ test subset. These numbers should be regarded as representative rather than as a fully converged estimate of the
population-level reconstruction accuracy. For the reionization history itself, the network achieves a small absolute reconstruction error (MAE = 0.0120 and RMSE = 0.0139), indicating that the predicted $x_{\mathrm{HI}}(z)$ tracks the target curves closely across the transition regime.

The signed residual distribution shows a mild negative bias ($\mu=-0.0069$; median $=-0.0028$) with a modest scatter ($\sigma=0.0176$), implying a slight tendency to overestimate the neutral fraction on average under the adopted sign convention. When propagated to summary parameters, the midpoint redshift is recovered more accurately than the duration in terms of dimensionless relative errors: the error statistics for $z_{50}$ are $\mathrm{MAE}=0.0046$ and $\mathrm{RMSE}=0.0100$, with a small positive bias ($\mu=0.0034$, median $=0.0015$) and a tight dispersion ($\sigma=0.0094$). In contrast, the duration $\Delta z$ is more challenging, with larger relative errors ($\mathrm{MAE}=0.0302$, $\mathrm{RMSE}=0.0378$) and a broader spread ($\sigma=0.0378$), although the mean bias remains close to zero ($\mu=-0.0019$; median $=-0.0012$). This asymmetry between midpoint and duration is the main scientific message of the representative noisy test: within the restricted prior, fixed-$k$ power-spectrum evolution is a stronger estimator of reionization timing than of the detailed width of the transition.

Overall, these representative noisy-input results support a cautious conclusion. Within the adopted model family and the favorable signal-to-noise regime of the idealized foreground-free noise model considered here, the inverse mapping from $\Delta^2_{21}(k,z)$ to $x_{\mathrm{HI}}(z)$ appears reasonably stable, especially for midpoint-related quantities. Duration estimates are less precise and are therefore more sensitive to both noise and prior assumptions.

\section{SUMMARY \& DISCUSSION}

We have investigated whether fixed-$k$ trajectories of the dimensionless 21 cm power spectrum, $\Delta^2_{21}(k,z)$, can be used to reconstruct the reionization history $x_{\mathrm{HI}}(z)$ within a restricted, physically motivated three-parameter model by 21cmFAST. The main result of this work is that, within the adopted three-parameter 21cmFAST prior, fixed-$k$ trajectories of $\Delta^2_{21}(k,z)$ contain enough information to reconstruct the reionization history $x_{\mathrm{HI}}(z)$ with a compact ANN. In the representative tests presented here, the network recovers the timing of reionization more reliably than the detailed width of the transition. This conclusion should be interpreted within the scope of the present setup, where the network serves as a prior-dependent neural-network-based reconstructor rather than a model-independent framework.

Within the present setup, the lower-$k$ trajectories provide the clearest connection to the global evolution of reionization. The reconstructions at $k=0.1$ and $0.5~h\,\mathrm{Mpc}^{-1}$ are broadly consistent with the expectation that larger-scale modes trace the global ionization trend more directly. The reconstructions at $k=1.0~h\,\mathrm{Mpc}^{-1}$ remain reasonably good within the present setup. We interpret this as a consequence of the restricted training family adopted here, in which the allowed histories occupy a relatively narrow manifold of smooth curves. The weak scale dependence seen in the present calculations should therefore be understood as a prior-dependent feature of the adopted model space; in a broader astrophysical parameter space, a stronger scale dependence may emerge.

The most informative result comes from the transition-interval analysis. In the representative one-split test reported here, the midpoint $z_{50}$ is recovered substantially more accurately than the duration $\Delta z$. This indicates that fixed-$k$ power-spectrum trajectories retain stronger information about the timing of reionization than about the detailed width of the transition, at least within the adopted prior. This hierarchy has a natural physical interpretation: the redshift at which the fixed-$k$ power-spectrum trajectory changes most rapidly is closely tied to the epoch when ionized structures of the corresponding scale grow and overlap, whereas the width of the transition depends more strongly on the detailed source history, bubble-merger process, and thermal assumptions. In that sense, the present calculation is best interpreted as an information-structure result for a restricted inverse problem.

The noise tests lead to a similarly qualified conclusion. In the favorable regime of the idealized foreground-free SKA1-Low-like thermal-plus-sample-variance model considered here, the reconstruction remains reasonably stable after noise is added, especially for midpoint-related quantities. This should not be interpreted as universal robustness. The network is trained on a restricted simulation prior, and its output should therefore be interpreted as reliable only within the corresponding model family.

A second important limitation is the small size of the training set relative to the number of ANN parameters. Although the histories in the present restricted prior form a smooth and low-dimensional family, training a 25057 parameter network on fewer than 200 training examples per fixed-$k$ split is intrinsically susceptible to overfitting. We have used an independent validation set, early stopping, leakage-safe preprocessing, a held-out test set, and the hidden-size ablation in Table~\ref{tab:hidden_ablation} to make the demonstration internally controlled. The ablation indicates that the qualitative recovery is not an obvious artifact of using only the 128-neuron hidden layer, but these safeguards do not by themselves establish robust extrapolation. The performance values reported here should therefore be read as an in-prior proof-of-concept benchmark. A larger and more densely sampled simulation suite is an essential next step: increasing the number of independent 21cmFAST realizations would allow repeated train/validation/test splits, more reliable variance estimates, explicit tests of sample-size convergence, and more meaningful comparisons with lower-capacity and non-neural baseline models. We therefore regard the expansion of the training set as a key requirement for future work before the approach can be used as a general inference tool.

Several steps are required before this approach can mature into a more 
broadly useful reconstruction tool. A key next step is to test which aspects 
of the present conclusion survive once the current prior compression is relaxed. 
Specifically, a major simplification in our current proof-of-concept setup 
is keeping the X-ray heating parameters fixed at their default values while restricting 
the three astrophysical parameters to the narrow ranges in Table 1. In a more 
realistic cosmic scenario, varying X-ray heating parameters (such as the X-ray 
efficiency $f_{\rm X}$ and the minimum spectral energy $E_0$) would introduce 
complex heating-ionization degeneracies into the 21~cm power spectrum, as the 
observable signal responds simultaneously to the thermal and ionization states of the IGM.

In particular, it will be important to determine whether the stronger recoverability of $z_{50}$ relative to $\Delta z$ persists beyond the present restricted model family when such heating-ionization degeneracies and broader parameter spaces are present. The training set should therefore be widened to cover a broader range of reionization histories, X-ray heating models, and source prescriptions. In addition, uncertainty quantification should be incorporated, so that the network can report not only a best-fit history but also a calibrated level of confidence and, ideally, a warning when an input trajectory lies outside its reliable domain of validity.

Observational realism must also be improved by propagating foreground residuals, 
calibration errors, mode loss, and inter-redshift covariance through end-to-end mock analyses. Such tests will clarify whether fixed-$k$ trajectories remain sufficient in realistic applications to break the added degeneracies, or whether information from multiple $k$-modes or more flexible compressed summaries will be required. Finally, reconstructed histories should be confronted with external probes, including CMB optical-depth constraints, Ly$\alpha$-based measurements, and high-redshift galaxy observations, so that the network output can be tested in a genuinely multi-probe framework rather than treated as a stand-alone result.

\begin{acknowledgements}
This work is supported by the National SKA Program of China (No.2020SKA0110401), NSFC (Grant No.~12103044), and Yunnan Provincial Key Laboratory of Survey Science with project No. 202449CE340002. 
\end{acknowledgements}

\appendix
\section{Comparison with a local linear interpolation baseline}

To test whether the ANN performance could be matched by a simpler non-neural method, we added a deterministic local linear interpolation baseline inspired by the referee's suggestion and by the broader goal of source-model-independent 21 cm power-spectrum interpretation discussed by \citet{2024A&A...687A.252G}. We do not attempt to reproduce the full method of Ghara et al. (2024); rather, we use a simple local interpolation baseline motivated by the referee's suggestion to test whether a non-neural local method can match the ANN performance within our restricted dataset. For each held-out test trajectory, we computed Euclidean distances in the same scaled 97-dimensional power-spectrum trajectory space used by the ANN, selected the nearest training trajectories, and predicted the neutral-fraction history as an inverse-distance-weighted linear combination of their target histories. We tested 1, 2, 4, 8, and 16 nearest neighbours; the two-neighbour case gave the best overall baseline performance. The comparison below uses the same held-out test subset and the same metric definitions as Table~\ref{tab:performance}.

\begin{table}[!htbp]
\begin{center}
\caption[]{Comparison between the ANN and the best local linear interpolation baseline on the same held-out test subset. For $x_{\rm HI}$, MAE and RMSE are absolute neutral-fraction errors over the transition interval. For $z_{50}$ and $\Delta z$, the entries are dimensionless relative errors defined by \mbox{Equations~\ref{eq:errz50} and~\ref{eq:errdz}}.} \label{tab:interp_baseline}
\begin{tabular}{lcccccc}
\hline\noalign{\smallskip}
Method & $x_{\rm HI}$ MAE & $x_{\rm HI}$ RMSE & $z_{50}$ MAE & $z_{50}$ RMSE & $\Delta z$ MAE & $\Delta z$ RMSE \\
\hline\noalign{\smallskip}
ANN & 0.0120 & 0.0139 & 0.0046 & 0.0100 & 0.0302 & 0.0378 \\
Local linear interpolation & 0.0350 & 0.0411 & 0.0101 & 0.0185 & 0.0870 & 0.1254 \\
\noalign{\smallskip}\hline
\end{tabular}
\end{center}
\end{table}

The interpolation baseline captures the broad smoothness of the restricted training family, but it does not match the ANN. Its MAE is larger by factors of approximately 2.9 for $x_{\rm HI}$, 2.2 for $z_{50}$, and 2.9 for $\Delta z$, while its RMSE is larger by factors of approximately 3.0, 1.9, and 3.3, respectively. This comparison supports the view that the ANN is using information from the full redshift-dependent trajectory beyond nearest-neighbour interpolation in the compressed dataset. At the same time, the baseline performs non-trivially well, reinforcing our interpretation that the present experiment is an in-prior proof-of-concept on a smooth, restricted model manifold rather than a model-independent reconstruction test.

\bibliographystyle{raa}
\bibliography{ms2026-0348}

\end{document}